\documentclass[twocolumn,showpacs,preprintnumbers,amsmath,amssymb]{revtex4}
\usepackage{amsfonts}
\usepackage{amsmath}
\usepackage{graphicx}
\usepackage{dcolumn}
\usepackage{bm}
\usepackage{overpic}
\usepackage{booktabs}

\begin{document}
\title{Information filtering via preferential diffusion}
\author{Linyuan L\"{u}}\email{linyuan.lue@unifr.ch}
\author{Weiping Liu}

\affiliation{$^1$Web Sciences Center, University of Electronic Science and Technology of China, Chengdu 611731, P.R. China\\$^2$Department of Physics, University of Fribourg, Chemin du Mus\'{e}e 3, CH-1700 Fribourg, Switzerland}

\date{\today}

\begin{abstract}
Recommender systems have shown great potential to address
information overload problem, namely to help users in finding
interesting and relevant objects within a huge information space.
Some physical dynamics, including heat conduction process and mass
or energy diffusion on networks, have recently found applications in
personalized recommendation. Most of the previous studies focus
overwhelmingly on recommendation accuracy as the only important
factor, while overlook the significance of diversity and novelty
which indeed provide the vitality of the system. In this paper, we
propose a recommendation algorithm based on the preferential
diffusion process on user-object bipartite network. Numerical
analyses on two benchmark datasets, \emph{MovieLens} and
\emph{Netflix}, indicate that our method outperforms the
state-of-the-art methods. Specifically, it can not only provide more
accurate recommendations, but also generate more diverse and novel
recommendations by accurately recommending unpopular objects.

\end{abstract}

\keywords{}

\pacs{89.75.Hc, 89.20.Ff, 05.70.Ln} 

\maketitle

\section{Introduction.}
The development of information technology brings great impact on
human society. Therein, the most significant aspect is the
revolutionary change in the ways of life. Twenty years ago, if one
wants to buy something, he/she has to personally go to a physical
shop and purchase, and then bring the things back home. It is
impossible for him/her to compare the commodities in different
markets located at different places in a short time. Now with the
growth of the Internet and World Wide Web, we can almost manage our
life at home. When we want to buy a book, we don't need to go to
bookstores any more to find it on bookshelves one by one, instead
what we need to do is typing the title of this book on the website
of Amazon --- an online retailer of books. If we want to buy a cell
phone, we can compare the prices on different web-shops at the same
time without any transportation fee. Formerly, we usually go to a
bar after working and enjoy making friends there, now we prefer
online dating that allows us to reach people over the world. In a
word, the Internet benefits us by providing a much more convenient
way to get what we want. However, as a coin has two sides, Internet
also brings us confusion
--- we face information overload. As we know, not all the
online information are good or true or favorable by surfers.
Therefore, we need to distinguish and select between valuable
information and junks. In this sense, to get what we want or the
most satisfied things becomes more and more difficult, since we face
much more choices than before. A useful information filtering
technology is search engine \cite{Brin1998,Kleinberg1999}, by which
users can find the relevant information with properly chosen
keywords or tags. However, search engines have two disadvantages
which limit their applications. Firstly, they lack the consideration
of personalization and thus return the same results to people no
matter what their preferences are. Secondly, search engines require
the users to know exactly what they want and extract some proper
keywords to do the searching. However, sometimes the tastes or
preferences can not be easily expressed by keywords or the users
don't even know what they want at all. In these cases, the search
engines are of no avail.

To address these problems, recommender systems rise in response to
the proper time and conditions, which form or work from a specific
type of information filtering technique that attempts to recommend
information items, such as movies, TV programs, videos, music,
books, news, images and web pages, that are likely to be of interest
to the users. The recommender systems don't require specified
keywords provided by users, instead they use the users' historical
activities and possible personal profiles to uncover their
preferences or potential interests. Many recommendation algorithms
have been developed, including collaborative filtering (CF)
\cite{Goldberg1992,Schafer2007,Shang2009}, content-based analysis
\cite{Pazzani2007}, spectral analysis \cite{Goldberg2001,Maslov2001}
and iterative self-consistent refinement \cite{Laureti2006,Ren2008}.
What most have in common is that they are based on similarity,
either of users or objects or both. Such approach is under high risk
of providing poor coverage of the space of relevant items. As a
result, with recommendations based on similarity rather than
difference, more and more users will be exposed to a narrowing band
of popular objects. Although it seems more accurate to recommend
popular objects than niche ones, being accurate is not enough
\cite{McNee2006}. It was pointed out that the recommendations that
are most accurate are sometimes not the recommendations that are
useful to users. For example, would you use such a system that
recommends the movies you indeed like but have seen before or just
watched in the cinema? Diversity and novelty are also important
criteria of algorithmic performance. A possible way to increase the
recommendation diversity is utilizing the tags of objects
\cite{Cattuto2007,ZhangZK,ZhangZK2}. Another promising way is
considering the dissimilar users' contribution. It was shown that
under the framework of collaborative filtering the dissimilar users
can contribute to both accuracy and diversity of personalized
recommendation \cite{Zeng2010}. However, these improvements are very
limited.

Recently, some physical dynamics, including mass diffusion
\cite{Zhou2007,Zhang2007b} and heat conduction process
\cite{Zhang2007a} have been applied to design recommender systems.
Zhou \emph{et al.} proposed a network-based inference method (NBI)
by considering the three-step mass diffusion starting from the
target user on a user-object bipartite network \cite{Zhou2007}. This
method has been demonstrated to be more accurate than the classical
CF algorithm while with lower computational complexity. However, it
has difficulty in generating diverse recommendations. The heat
conduction process has been found its effectiveness in providing a
diverse recommendation at the cost of accuracy. This
diversity-accuracy dilemma can be effectively solved by coupling
these two processes \cite{ZhouPNAS2010}. It was shown that not only
does the hybrid algorithm outperform other methods but that, without
relying on any semantic or context-specific information, it can be
tuned to obtain significant gains in both accuracy and diversity of
recommendations.

With the same motivation, we proposed an algorithm based on a
preferential mass diffusion process on user-object bipartite
networks, without consideration of heat conduction which may
stealthily hurt accuracy. Numerical analyses on two benchmark
datasets show that our method can give higher accurate as well as
more diverse and novel recommendations than the hybrid algorithm,
because of its high accurate recommendations on low-degree objects.

\section{Preferential diffusion method}
A recommender system can be represented by a bipartite network $G(U,
O, E)$, where $U=\{u_{1},u_{2},...,u_{m}\}$,
$O=\{o_{1},o_{2},...,o_{n}\}$ and $E=\{e_{1},e_{2},...,e_{q}\}$ are
the sets of users, objects and links respectively \cite{Shang2009}.
Denote by $A_{m\times{n}}$ the adjacency matrix, where the element
$a_{i\alpha}$ equals 1 if $u_i$ has collected object $o_\alpha$, and
0 otherwise.

The essential task of a recommender system is to generate a ranking
list of the target user's uncollected objects. The original
diffusion-based recommendation algorithm, called network-based
inference (NBI), was proposed in Ref. \cite{Zhou2007}. It was
referred as ProbS algorithm in Ref. \cite{ZhouPNAS2010}. NBI works
by assigning objects an initial level of resource denoted by the
vector $\mathbf{f}$ (where $f_\alpha$ is the resource possessed by
object $o_\alpha$), and then redistributing it via the
transformation $\mathbf{f}'=W\mathbf{f}$, where
\begin{equation}
w_{\alpha\beta}=\frac{1}{k_{o_\beta}}\sum_{l=1}^{m}\frac{a_{l\alpha}a_{l\beta}}{k_{u_l}},\label{NBI}
\end{equation}
is the resource transfer matrix, and $k_{o_\beta}=\sum_{i=1}^n
a_{i\beta}$ and $k_{u_l}=\sum_{\gamma=1}^m a_{l\gamma}$ denote the
degrees of object $o_\beta$ and user $u_l$ respectively. For a target user $u_i$, we assign one unit
resource on those objects already collected by $u_i$ for simplicity,
thus the initial resource vector $\mathbf{f}$ can be written as
\begin{equation}
f_\alpha=a_{i\alpha}.\label{initial}
\end{equation}
That is to say, if object $o_\alpha$ is collected by user $u_i$ then
it has one unit resource, otherwise 0. With this initial resource vector, the result of NBI is equivalent to a three-step random walk process starting from the target user on a bipartite network \cite{LiuWP2010}. Note that, if the initial resource vector is normalized by the target user's degree, namely $f_\alpha=a_{i\alpha}/k_{ui}$, the results are exactly the same. In fact, the process of NBI is equivalent to resource-allocation which is also a random-work-based process. Given the initial resource distribution as shown in Eq.~\ref{initial}, the resource of each object will be redistributed according to Eq.\ref{NBI} where $w_{\alpha\beta}$ indicates how many proportion of resource that object $\alpha$ gives to object $\beta$. Then after the
resource-allocation process, we obtain the final
resource possessed by each object by summing up all the resources distributed from other objects. The recommendation list for user $u_i$ is generated by ranking all his/her uncollected objects in
decreasing order according to their final resource.

A heterogenous initial resource distribution NBI algorithm
(abbreviate as Heter-NBI) was proposed by Zhou \emph{et al.}
\cite{Zhou2008}, where the initial resource of object $o_\alpha$ is
proportional to $k_{o_\alpha}^\theta$.Thus the initial resource vector of Heter-NBI can be written as $f_\alpha=a_{i\alpha}\cdot k_{o\alpha}^\theta$ where $\theta$ is a negative parameter. It was shown that Heter-NBI
can give more accurate recommendations than the standard NBI. There
are other two advanced recommendation algorithms. One is an improved
algorithm by eliminating redundant correlations (called RE-NBI for
short) \cite{ZhouNJP2009}, which is defined as
\begin{equation}
\mathbf{f}'=(W+\eta{W^2})\mathbf{f},
\end{equation}
where the elements of matrix $W$ are defined by Eq.~\ref{NBI}, the
initial resource vector $\mathbf{f}$ is defined by Eq. \ref{initial}
and $\eta$ is a free parameter. This method has been approved to
outperform some classical methods, such as the global ranking
method, the cosine-similarity-based collaborative filtering
\cite{Herlocker2004}, NBI and Heter-NBI for both accuracy and
diversity by considering the high-order correlations between
objects. The other method, referred as Hybrid-PH in this paper, is
proposed in Ref. \cite{ZhouPNAS2010}, which is a hybrid algorithm
combining the HeatS (i.e., heat conduction) and ProbS (i.e., mass
diffusion) by incorporating the hybridization parameter $\lambda$
into the transition matrix normalization:
\begin{equation}
w_{\alpha\beta}=\frac{1}{k_{o_\alpha}^{1-\lambda}k_{o_\beta}^\lambda}\sum_{l=1}^m\frac{a_{l\alpha}
a_{l\beta}}{k_{u_l}},\label{Hybrid}
\end{equation}
where $\lambda=0$ gives the pure HeatS algorithm, and $\lambda=1$
gives the ProbS (i.e., NBI).

Based on mass diffusion method and motivated by enhancing the
algorithm's ability to find unpopular and niche objects, we propose
a preferential diffusion (PD) method for recommendation in
user-object bipartite networks. The basic idea is that at the last
step (i.e., diffusing from users to objects), the amount of resource
that an object $o_\alpha$ received is proportional to
$k_{o_\alpha}^\varepsilon$, where $\varepsilon\leq 0$ is a free
parameter. 
In this case, the resource transfer matrix reads
\begin{equation}
w_{\alpha\beta}=\frac{1}{k_{o_\beta}\cdot{k_{o_\alpha}^{-\varepsilon}}}\sum_{l=1}^{m}\frac{a_{l\alpha}a_{l\beta}}{\mathcal
{M}},\label{PD-C}
\end{equation}
where $\mathcal
{M}=\sum_{r=1}^n{a_{lr}k_{o_r}^\varepsilon}=k_{ul}\cdot
\mathrm{E}(a_{lr}k_{o_r}^\varepsilon)$.
$\mathrm{E}(a_{lr}k_{o_r}^\varepsilon)$ indicates the mean value of
$k_{o_r}^\varepsilon$ over all the objects having been collected by
user $u_l$. Here we consider the simplest initial resource vector
defined by Eq. \ref{initial}. Clearly, when $\varepsilon=0$ it will
degenerate to NBI. Notice that, if we consider the NBI algorithm as
a three step diffusion starting from target user to final objects
(i.e., user$\rightarrow$object$\rightarrow$user$\rightarrow$object),
then the Heter-NBI algorithm is essentially equivalent to the
algorithm with preferential diffusion only at the first step, while
PD considers the third step. However, their motivations are
essentially different. Heter-NBI emphasizes that users who
co-collected unpopular objects are more similar to each other than
those co-collected popular objects. And thus the target user
distributes more resource to his/her more similar users by giving
more resource to their co-collected unpopular objects. However,
after the third step diffusion the resource still can be centralized
on some popular objects. The PD algorithm directly punishes the
popular object by assigning more resource to the low-degree objects
at the last step. Experimental results show that considering the
preferential diffusion at the last step is much more effective than
at the first step. In order to show that preferential diffusion at
first step (i.e., Heter-NBI) and at last step (i.e., PD) play
different roles in recommendation, we further investigate the PD
algorithm with heterogenous initial resource distribution, called
Heter-PD, which is controlled by two tunable parameters. Comparing
with all the mentioned algorithms in this paper, Heter-PD performs
the best over all five evaluation metrics considered in this paper
(see section 3 for the definitions of evaluation metrics). Comparing
Eq.~\ref{Hybrid} with Eq.~\ref{PD-C}, we can find that if we assume
that for user $u_l$ who has collected object $o_\beta$, the
approximation $\mathrm{E}(a_{lr}k_{o_r}^\varepsilon)\approx
k_{o_\beta}^\varepsilon$ holds, namely the mean value of
$k_{o_r}^\varepsilon$ over all the objects having collected by user
$u_l$ always equals $k_{o_\beta}^\varepsilon$, PD is equivalent to
the hybrid algorithm by setting $\varepsilon=\lambda-1$. However,
this assumption is too strong to be satisfied in reality.

Note that, we didn't consider the preferential diffusion at the
second step from the object side to the user side (PD-II for short).
The main reason is that this method may lead to some illogical
results. Considering the case that the target user $u_i$ selected a
very popular object $o_\alpha$ which is also selected by another
user $u_j$ who is assumed to be a new user of the system and only
selected $o_\alpha$. Via the PD-II method, $u_j$ will obtain more
resource from $o_\alpha$ than other users who also selected
$o_\alpha$, leading to the conclusion that $u_j$ is more similar to
$u_i$. Apparently this result is wrong, since a new user usually
selects popular objects, which is a common behavior in such kind of
systems \cite{Shang2010}, and it is unreasonable to say this new
user is more similar to the target user just according to such a
common behavior. In addition, we have tested the performance of
PD-II method. Comparing with standard NBI method, the improvement of
accuracy (measured by ranking score) is very slight around 1\% on
MovieLens data and 0.6\% on Netflix data. Therefore, we didn't
consider this method for further analysis.

\section{Data and metrics}
To test the algorithmic performance, we use two benchmark datasets.
The \emph{MovieLens} (http://www.grouplens.org/) data consists of
1682 movies (objects) and 943 users who can vote for movies with
five level ratings from 1 (i.e., worst) to 5 (i.e., best). The
original data contains $10^5$ ratings. Here we only consider the
ratings higher than 2. After coarse gaining the data contains 82520
user-object pairs. The \emph{Netflix} data
(http://www.netflixprize.com/) is a random sampling of the whole
records of user activities in \emph{Netflix.com}. It consists of
10000 users, 6000 movies and 824802 links. Similar to the
\emph{MovieLens} data, only the links with ratings no less than 3
are considered. After data filtering, there are 701947 links left.
To test the algorithmic performance, the data (i.e., known links) is
randomly divided into two parts: The training set $E^T$ contains
90\% of the data, and the remaining 10\% of data constitutes the
probe set $E^P$. Notice that, any isolate object can not be
recommended to users through the algorithms considered in this
paper. Therefore to ensure the connectivity of the whole network,
each time before moving a link to the probe set, we first check if
this removal will result in isolate user or object, and we do not
allow the removal that leads to unconnected nodes.

Accuracy is the most important aspect to evaluate the recommendation
algorithmic performance. A good algorithm is expected to give
accurate recommendations, namely higher ability to find what the
users like. Here we use \emph{Ranking Score} \cite{Zhou2007} to
measure the ability of a recommendation algorithm to produce a good
ordering of objects that matches the user's preference. For a target
user, the recommender system will return a ranking list of all his
uncollected object to him. For each hidden user-object relation
(i.e., the link in probe set), we measure the rank of this object in
the recommendation list of this user. For example, if there are 1000
uncollected objects for user $u_i$, and object $o_\alpha$ is at 10th
place, we say the position of this object is 10/1000, denoted by
$RS_{i\alpha} =0.01$. A good algorithm is expected to give high
ranks to the hidden objects, and thus leading to small $RS$.
Averaging over all the hidden user-object relations, we obtain the
mean value of ranking score $RS$ that can be used to evaluate the
algorithm's accuracy, namely
\begin{equation}
RS=\frac{1}{|E^P|}\sum_{i\alpha \in E^P}RS_{i\alpha},
\end{equation}
where $i\alpha$ denotes the probe link connecting $u_i$ and
$o_\alpha$. Clearly, the smaller the ranking score, the higher the
algorithm's accuracy, and vice versa. Since real users usually
consider only the top part of the recommendation list, a more
practical measure may be to consider the number of user's hidden
links contained in the top-$L$ places. Therefore, we adopt another
accuracy metric called \emph{Precision}. For a target user $u_i$,
the precision of recommendation, $P_i(L)$, is defined as
\begin{equation}
P_i(L) = \frac{d_{i}(L)}{L},
\end{equation}
where $d_i(L)$ indicates the number of relevant objects (namely the
objects collected by $u_i$ in the probe set) in the top-$L$ places
of recommendation list. Averaging the individual precisions over all
users with at least one hidden link, we obtain the mean precision
$P(L)$ of the whole system.

Besides accuracy, diversity is taken into account as another
important aspect to evaluate the recommendation algorithm. There are
two kinds of diversity. One is called \emph{Inter-Diversity} which
considers the uniqueness of different users' recommendation lists.
Given two users $u_i$ and $u_j$, the difference between their
recommendation lists can be measured by the Hamming distance
\cite{Zhou2008},
\begin{equation}
H_{ij}(L) = 1 - \frac{C_{ij}(L)}{L},
\end{equation}
where $C_{ij}(L)$ is the number of common objects in the top-$L$
places of both lists. Clearly, if $u_i$ and $u_j$ have the same
list, $H_{ij}(L)=0$, while if their lists are completely different,
$H_{ij}(L)=1$. Averaging $H_{ij}(L)$ over all pairs of users we
obtain the mean distance $H(L)$, for which greater or lesser values
mean, respectively, greater or lesser personalization of users'
recommendation lists. A good algorithm should not only give diverse
recommendations among users (i.e., high inter-diversity), but also
provide diverse recommendations for a single user (i.e., high
intra-diversity) \cite{ZhouNJP2009,Ziegler2005}. The latter can be
measured by \emph{Intra-Similarity}. For a target user $u_i$, his
recommended objects are \{$o_1,o_2,\cdots,o_L$\}, then the
intra-similarity of $u_i$'s recommendation list is defined as
\cite{ZhouNJP2009}:
\begin{equation}
I_i(L)=\frac{1}{L(L-1)}\sum_{\alpha\neq \beta}s^o_{\alpha\beta},
\end{equation}
where $s^o_{\alpha\beta}$ is the similarity between objects
$o_\alpha$ and $o_\beta$ in $u_i$'s recommendation list. There are
many similarity indices that can be used to quantify the similarity
between objects \cite{Zhou2009}. Here we adopt the widely used
cosine similarity to measure object similarity. For two objects
$o_\alpha$ and $o_\beta$ their similarity is defined as
\begin{equation}
s^o_{\alpha\beta}=\frac{1}{\sqrt{k_{o_\alpha}k_{o_\beta}}}\sum_{l=1}^m
a_{l\alpha}a_{l\beta}.
\end{equation}
Averaging $I_i(L)$ over all users we obtain the mean
intra-similarity $I(L)$ for the system. A good recommendation
algorithm is expected to give fruitful recommendations and has the
ability to guide or help the users to exploit their potential
interest fields, and thus leads to a lower intra-similarity (i.e.,
higher intra-diversity).

High accurate recommendations might not be satisfied by the users.
For example, recommending popular film \emph{Avatar} to a user on
\emph{MovieLens} website is not always the best, because he/she
might have already seen this film at the cinema. A diverse
recommender system is expected to find the niche or unpopular
objects that can not be easily known by other ways yet match users'
preferences. The metric \emph{Popularity} quantifies the capacity of
an algorithm to generate novel and unexpected results, that is to
say, to recommend less popular items unlikely to be already known
about. The simplest way to calculate popularity is to use the
average collected times over all the recommended items, as:
\begin{equation}
\label{novelty} N_i(L) = \frac{1}{L}\sum_{o_\alpha\in
O_{R}^{i}}{k_{o_\alpha}},
\end{equation}
where $O_{R}^{i}$ is the recommendation list for user $u_i$.
Clearly, lower popularity indicates higher novelty and surprisal.
Averaging $N_i(L)$ over all users we obtain the mean popularity
$N(L)$ for the system.


\begin{table*}
\caption{Algorithmic performance for \emph{MovieLens} data. The
precision, intra-similarity, hamming distance and popularity are
corresponding to $L=50$. Heter-NBI is an abbreviation of NBI with
heterogenous initial resource distribution, proposed in Ref.
\cite{Zhou2008}. RE-NBI is an abbreviation of redundant-eliminated
NBI, proposed in Ref. \cite{ZhouNJP2009}. Hybrid-PH refers to the
hybrid method which combines ProbS and HeatS algorithms. PD is an
abbreviation of preferential diffusion method presented in this
paper. Heter-PD is an abbreviation of PD with heterogenous initial
resource distribution. The parameters (ranging in the interval [0,1]
for Hybrid-PH and [-1,0] for the rest five algorithms with step
0.05) for the parameter-dependent algorithms are set as the ones
corresponding to the lowest ranking scores (for Heter-NBI,
$\theta_{\texttt{opt}}=-0.80$; for RE-NBI,
$\eta_{\texttt{opt}}=-0.75$; for Hybrid-PH,
$\lambda_{\texttt{opt}}=0.20$; for PD,
$\varepsilon_{\texttt{opt}}=-0.85$; for Heter-PD,
$(\theta,\varepsilon)_{\texttt{opt}}=(-0.25,-0.8)$). Each number is
obtained by averaging over five runs with independently random
division of training set and probe set. The entries corresponding to
the best performance over all methods (except Heter-PD) are
emphasized in black.}
\begin{center}
\begin{tabular} {ccccccc}
  \hline \hline
   Algorithms     & Ranking Score  &  Precision &  Intra-Similarity & Hamming Distance & Popularity  \\
   \hline
   NBI        & 0.106 & 0.071 & 0.355 & 0.617 & 233  \\
   Heter-NBI  & 0.101 & 0.074 & 0.340 & 0.680 & 220  \\
   RE-NBI     & \textbf{0.082} & \textbf{0.085} & 0.326 & 0.788 & 189  \\
   Hybrid-PH  & 0.085 & 0.083 & 0.296 & 0.821 & 167  \\
   PD         & \textbf{0.082} & 0.084 & \textbf{0.282} & \textbf{0.847} & \textbf{155}  \\
   \hline
   Heter-PD   & 0.081 & 0.086 & 0.278 & 0.858 & 153  \\
   \hline \hline
    \end{tabular}\label{table1}
\end{center}
\end{table*}

\begin{table*}
\caption{Algorithmic performance for \emph{Netflix} data. The
precision, intra-similarity, hamming distance and popularity are
corresponding to $L=50$. The parameters (ranging in the interval
[0,1] for Hybrid-PH and [-1,0] for the rest five algorithms with
step 0.05) for the parameter-dependent algorithms are set as the
ones corresponding to the lowest ranking scores (for Heter-NBI,
$\theta_{\texttt{opt}}=-0.70$; for RE-NBI,
$\eta_{\texttt{opt}}=-0.75$; for Hybrid-PH,
$\lambda_{\texttt{opt}}=0.20$; for PD,
$\varepsilon_{\texttt{opt}}=-0.85$; for Heter-PD,
$(\theta,\varepsilon)_{\texttt{opt}}=(-0.2,-0.8)$). Each number is
obtained by averaging over five runs with independently random
division of training set and probe set. The entries corresponding to
the best performance over all methods (except Heter-PD) are
emphasized in black.}
\begin{center}
\begin{tabular} {ccccccc}
  \hline \hline
   Algorithms     & Ranking Score  &  Precision &  Intra-Similarity & Hamming Distance & Popularity  \\
   \hline
   NBI        & 0.050 & 0.050 & 0.366 & 0.424 & 2366  \\
   Heter-NBI  & 0.047 & 0.051 & 0.341 & 0.545 & 2197  \\
   RE-NBI     & \textbf{0.039} & \textbf{0.062} & 0.336 & 0.629 & 2063  \\
   Hybrid-PH  & 0.045 & 0.057 & 0.311 & 0.625 & 1998  \\
   PD         & 0.041 & 0.057 & \textbf{0.295} & \textbf{0.639} & \textbf{1900}  \\
   \hline
   Heter-PD   & 0.040 & 0.057 & 0.266 & 0.708 & 1742  \\
   \hline \hline
    \end{tabular}\label{table2}
\end{center}
\end{table*}

\begin{figure}
\scalebox{0.75}[0.75]{\includegraphics{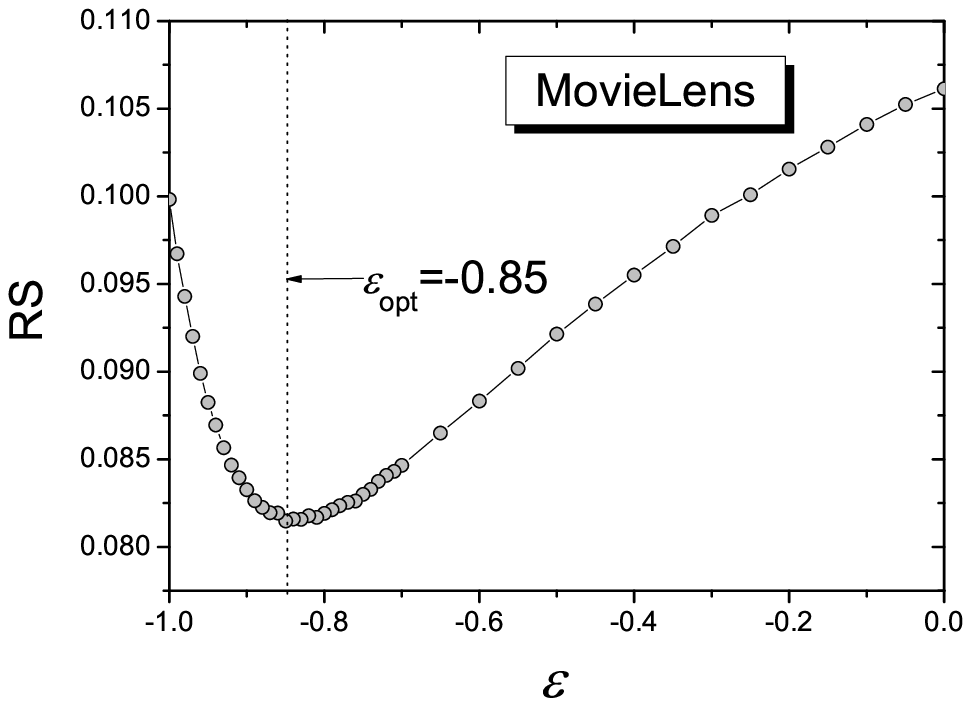}}
\scalebox{0.75}[0.75]{\includegraphics{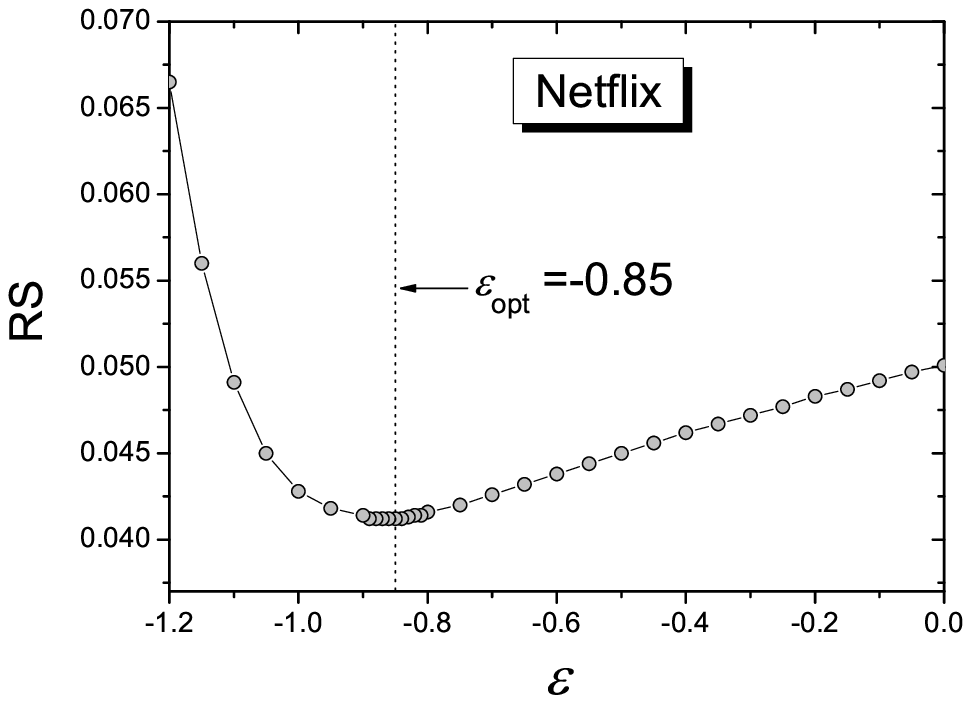}}\caption{The
ranking score $RS$ \emph{vs.} $\varepsilon$. Each data point is
obtained by averaging over five runs, each of which has an
independently random division of training set and probe set. The
optimal parameters $\varepsilon$ for \emph{MovieLens} and
\emph{Netflix}, corresponding to the minimal $RS$, both equal to
-0.85.}\label{RS_a}
\end{figure}

\begin{figure}
\scalebox{0.75}[0.75]{\includegraphics{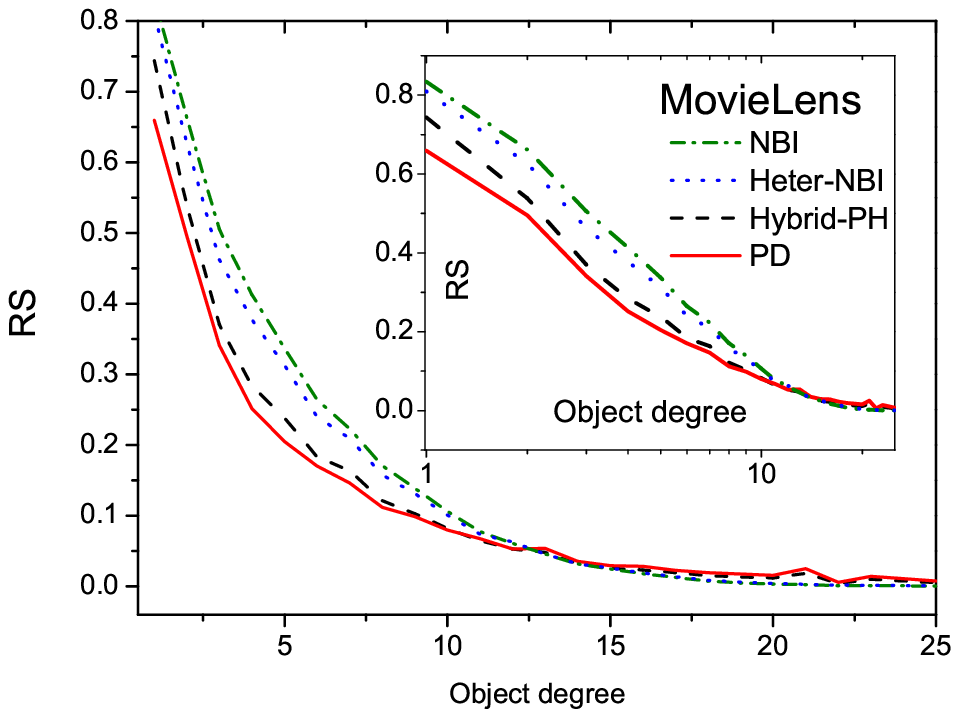}}
\scalebox{0.75}[0.75]{\includegraphics{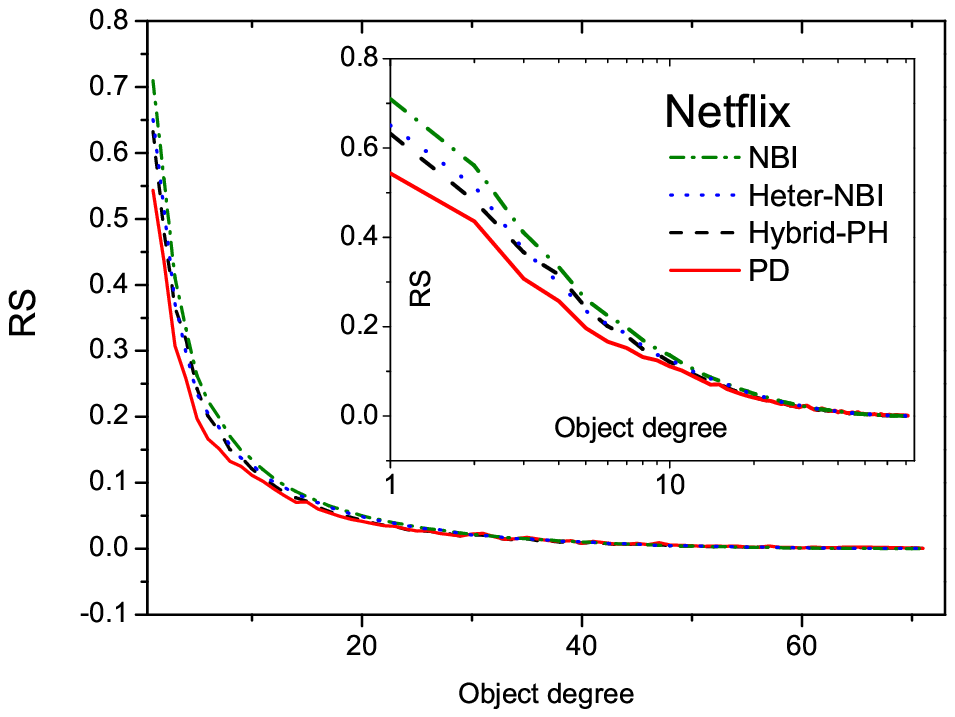}}\caption{(Color online) The
dependence of ranking score $\langle RS\rangle$ on the object
degree. For a given $x$, its corresponding \emph{RS} is obtained by
averaging over all the objects whose degrees are in the range of
($a(x^2-x)$,$a(x^2+x)$], where $a$ is chosen as $\frac{1}{2}\log5$
for a better illustration. Insets show \emph{RS} against logarithm
of $x$.}\label{Ko_RS}
\end{figure}

\begin{figure}
\scalebox{0.75}[0.75]{\includegraphics{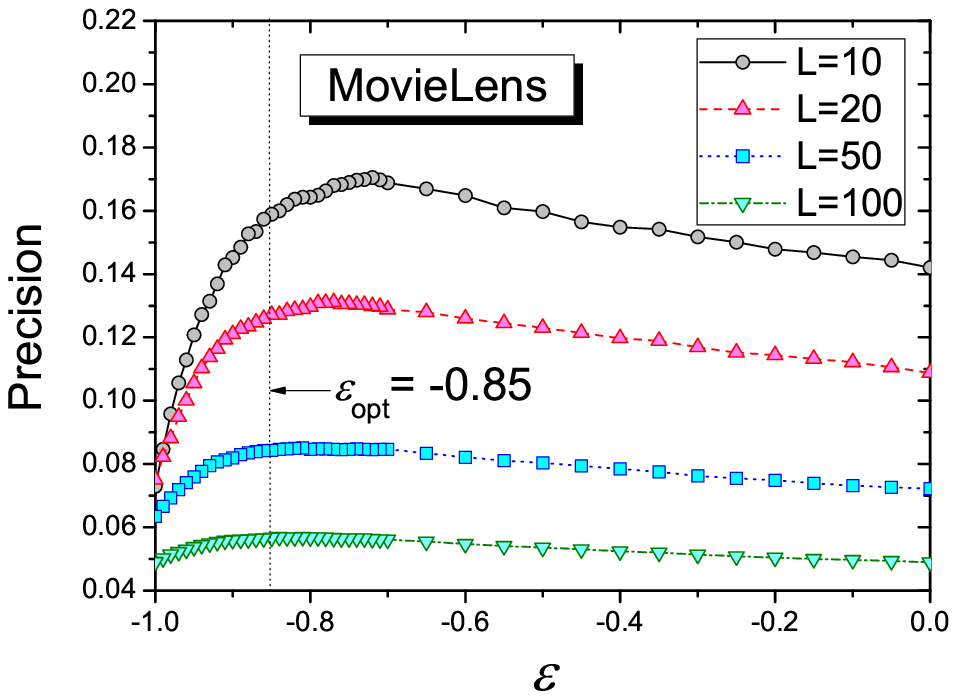}}
\scalebox{0.75}[0.75]{\includegraphics{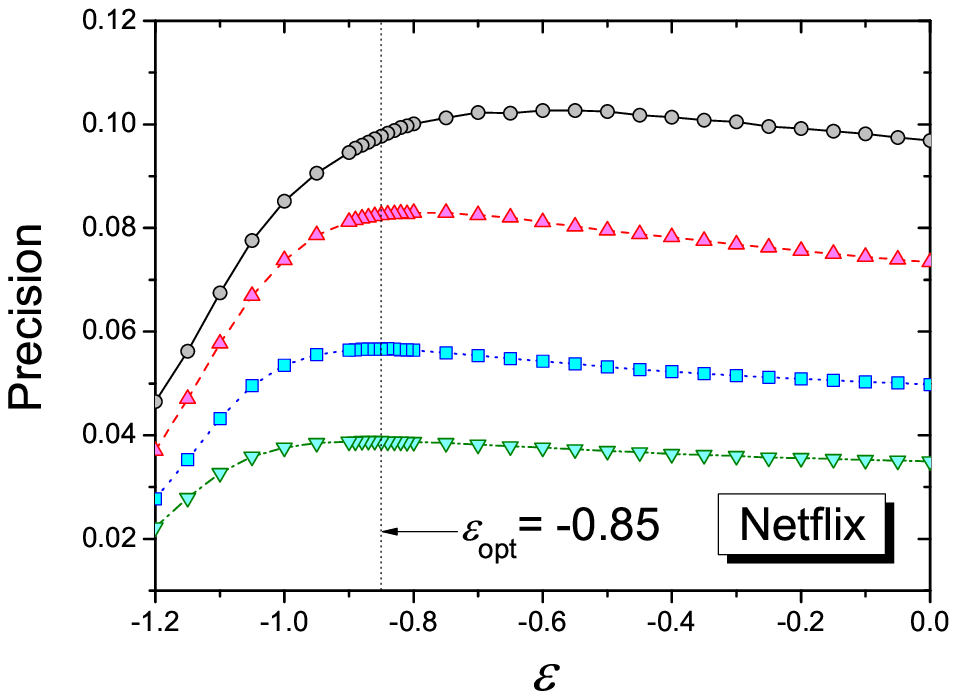}}\caption{(Color online) The
dependence of precision on parameter $\varepsilon$. Each data point
is obtained by averaging over five independent runs with data
division identical to the case shown in figure~\ref{RS_a}. The
vertical dotted line indicates the optimal parameter $\varepsilon$
subject to the lowest ranking score.}\label{Pre_a}
\end{figure}

\section{Results}
Summaries of the results for all algorithms and metrics on
\emph{MovieLens} and \emph{Netflix} datasets are shown respectively
in Table~\ref{table1} and Table~\ref{table2}. The so-called
\emph{optimal parameters} are subject to the lowest ranking score.
And the other four metrics, namely precision, intra-similarity,
hamming distance and popularity, are obtained at the optimal
parameters. Clearly, PD outperforms Heter-NBI over all the five
evaluation metrics. Among all four previous algorithms, Re-NBI gives
the highest accuracy by considering the high-order correlations
between objects, while Hybrid-PH has the best performance on
diversity and novelty. Comparing with these two outstanding
algorithms, PD can reach or closely near the best accuracy without
considering high-order correlation between objects, and provide much
more diverse results. By considering the heterogenous initial
resource distribution the algorithmic performance can be further
improved. For example, in \emph{MovieLens} Heter-PD decreases the
ranking score to 0.081 with the parameters $\theta=-0.25$ and
$\varepsilon=-0.8$, which is the lowest among all the methods
referred in this paper. Although with a heterogenous initial
resource distribution, both accuracy and diversity can be improved,
comparing with pure PD algorithm, such improvements are less
remarkable. This indicates that PD actually plays the main role of
improvements.

For PD algorithm, the dependence of parameter $\varepsilon$ on
accuracy measured by ranking score is shown in figure~\ref{RS_a}.
The optimal values of parameter $\varepsilon$ corresponding to the
lowest ranking score on two datasets are both equal to 0.85.
Comparing with the standard case NBI, namely $\varepsilon=0$, the
ranking score can be reduced by 23\% for \emph{MovieLens} and 18\%
for \emph{Netflix}. We further investigate the dependence of ranking
score (\emph{RS}) on the object degree of four methods, namely NBI,
Heter-NBI, Hybrid-PH and our method PD. The results are shown in
figure~\ref{Ko_RS}. Notice that, for a given $x$, its corresponding
\emph{RS} is obtained by averaging over all the objects whose
degrees are in the range of
($\frac{x^2-x}{2}\log5$,$\frac{x^2+x}{2}\log5$]. Insets show the
\emph{RS} against logarithm of $x$. It can be seen that the ranking
score decreases with the increasing of the object degree for all
these four algorithms. This indicates that in average popular
objects can be more accurately recommended than the unpopular
objects. The significant differences of these four algorithms are
embodied on their ability of accurately recommending unpopular
objects. Clearly, PD works best for this task, and is followed by
Hybrid-PH. Moreover, comparing the results of Heter-NBI with PD, we
can see that although they both consider the preferential diffusion
from user to object, considering at the first step (i.e., Heter-NBI)
has much less effect on the unpopular objects than directly acting
on the final step (i.e., PD).

Figure~\ref{Pre_a} shows how the precision changes with the
parameter $\varepsilon$ for four typical lengths of recommendation
list. Given $L$ there exists an optimal parameter $\varepsilon$
leading to the highest precision. Although this optimal parameter
$\varepsilon_1$ is different from that subject to the lowest ranking
score $\varepsilon_2$, the precision obtained with $\varepsilon_2$
is also considerably higher than that obtained by NBI. For example,
when $L=50$ with the optimal parameter corresponding to the lowest
ranking score, the precision is prominently improved by 18\% and
14\% for \emph{MovieLens} and \emph{Netflix} respectively.

\begin{figure}
\scalebox{0.75}[0.75]{\includegraphics{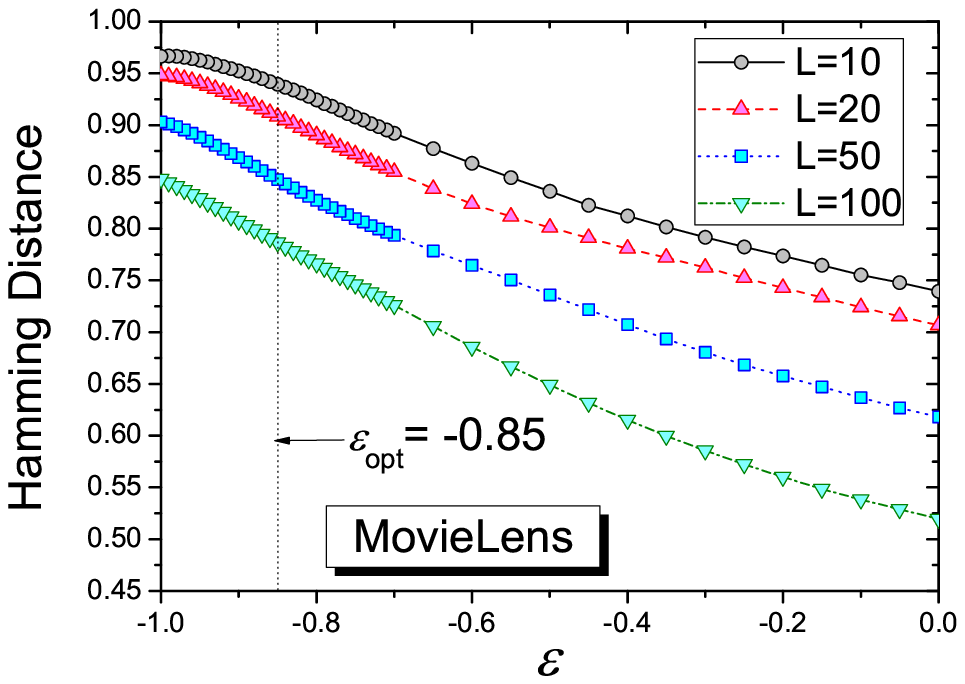}}
\scalebox{0.75}[0.75]{\includegraphics{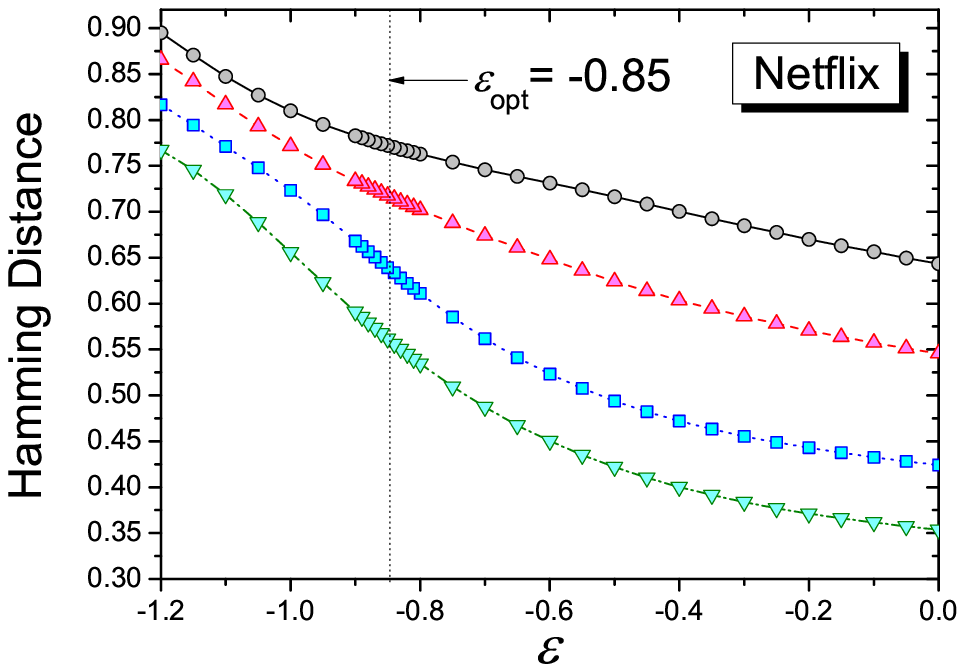}}\caption{(Color online) The
Hamming distance \emph{vs.} $\varepsilon$. Each data point is
obtained by averaging over five independent runs with data division
identical to the case shown in figure~\ref{RS_a}. The vertical
dotted line indicates the optimal parameter $\varepsilon$ subject to
the lowest ranking score.}\label{HD_a}
\end{figure}

\begin{figure*}
\scalebox{0.3}[0.3]{\includegraphics{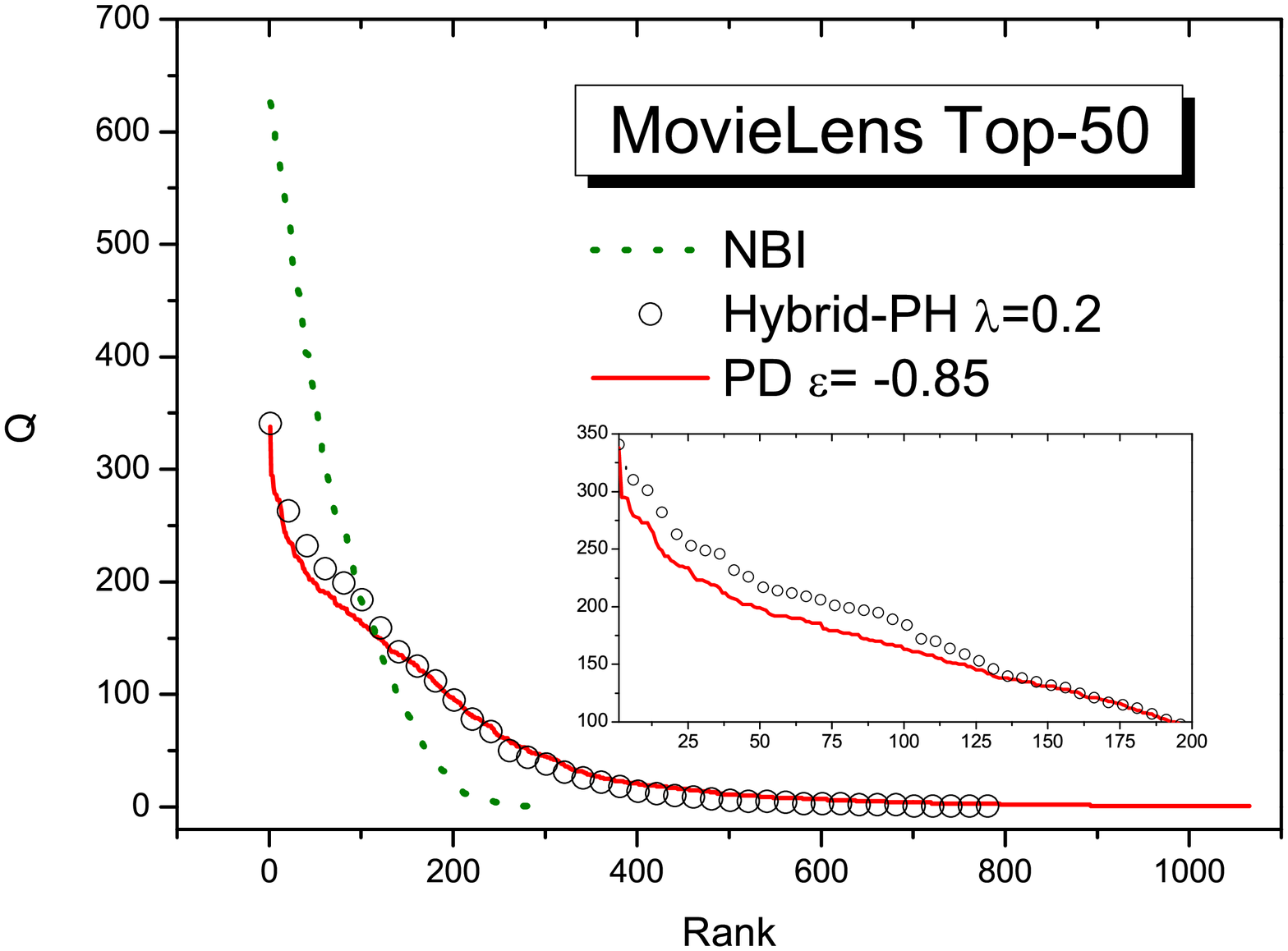}}
\scalebox{0.3}[0.3]{\includegraphics{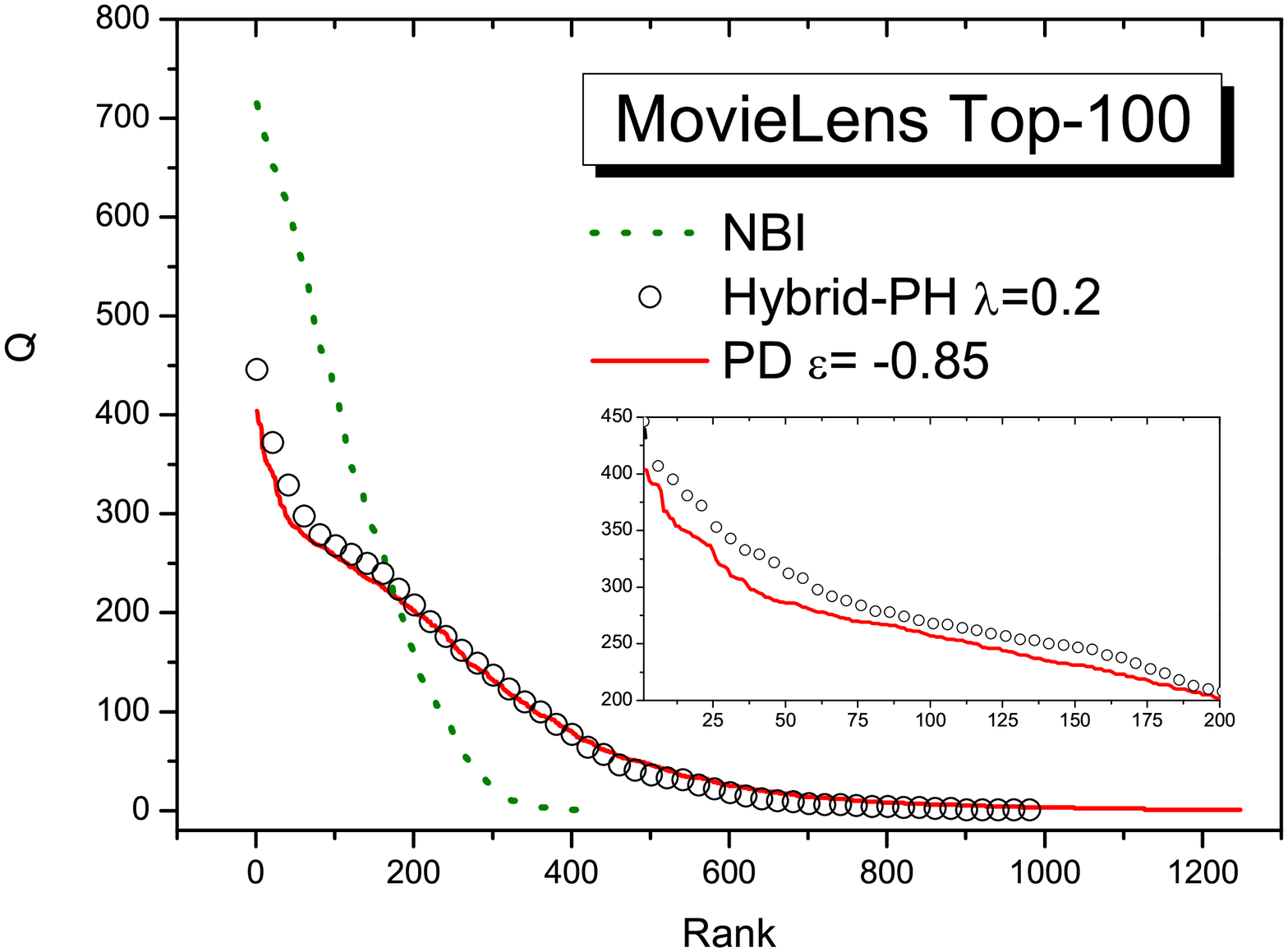}}
\scalebox{0.3}[0.3]{\includegraphics{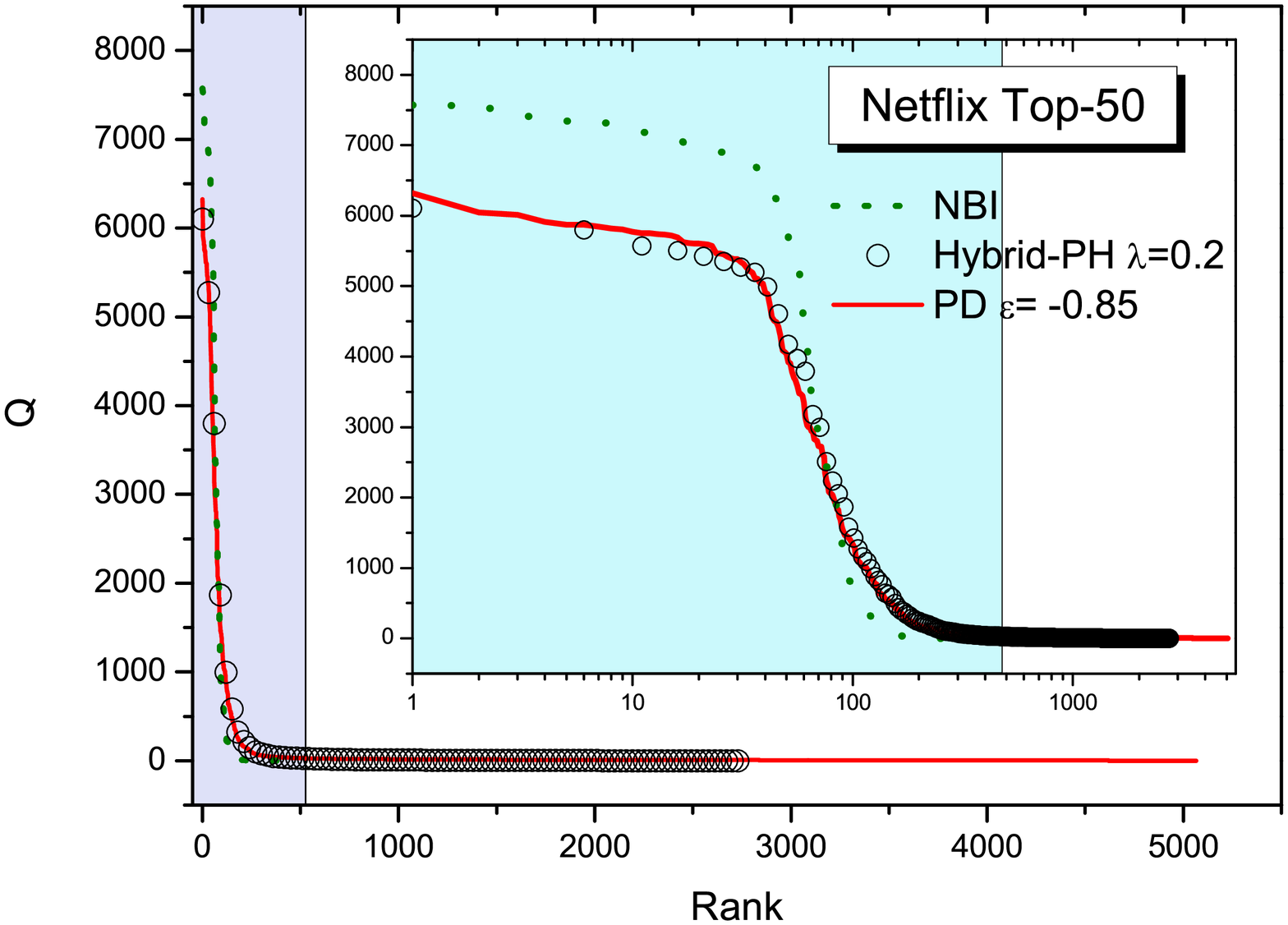}}
\scalebox{0.3}[0.3]{\includegraphics{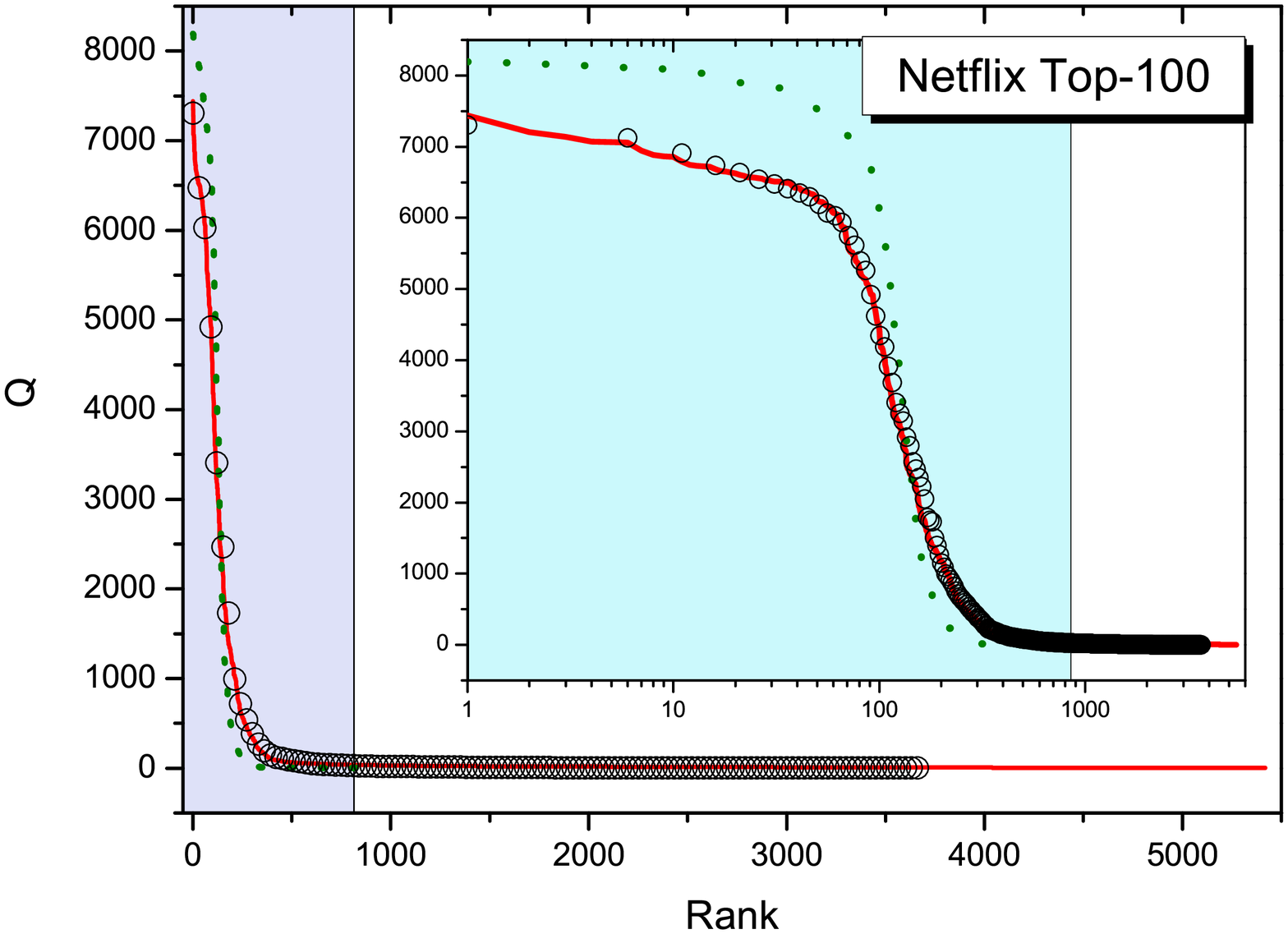}}\caption{(Color online) The
relationship between the recommended times $Q$ of objects and their
ranks for two datasets. Insets of two \emph{MovieLens} sub-figures
show the results of top-200 frequently recommended objects. Insets
of two \emph{Netflix} sub-figures show $Q$ against logarithm of $x$
(i.e., Rank). For NBI, only the objects inside the blue region have
the chance to be recommended.}\label{Q_a}
\end{figure*}

Hamming distance actually measures the ability that an algorithm
give personalized recommendation. How the parameter $\varepsilon$
affects the Hamming distance is shown in figure~\ref{HD_a}. Clearly,
a smaller $\varepsilon$ leads to a higher Hamming distance (i.e.,
higher inter-diversity), and thus a more personalized
recommendation. Comparing with the standard case NBI, given $L=50$,
Hamming distance can be enhanced by 37\% for \emph{MovieLens} and
56\% for \emph{Netflix} with optimal parameters corresponding to
their respective lowest ranking scores, even higher than the
Hybrid-PH algorithm. As a result, our method has higher ability to
find the niche (unpopular) objects that may be liked by users, and
thus give a more personalized recommendation to the target user. To
give more evidences, for a given algorithm we collect the top-$L$
recommended objects for each user. Denote by $d$ the number of
distinct objects among all the recommended objects. Then we rank the
$d$ objects according to their recommended times, denoting by $Q_i$
($i = 1$, $\cdots$, $d$), in decreasing order. The relationships
between the objects' recommended times $Q$ and their ranks are shown
in figure~\ref{Q_a}. We have tested for many different $L$, and here
take $L=50$ and $L=100$ as typical examples. Two important phenomena
can be obtained from figure~\ref{Q_a}. Firstly, comparing three
algorithms, NBI, Hybrid-PH with $\lambda=0.2$ and PD with
$\varepsilon=-0.85$, we have $d_{PD}>d_{Hybrid}>d_{NBI}$. That is to
say PD provides larger number of distinct objects to users than NBI
and Hybrid-PH. For example, when the length of recommendation list
is 50, in \emph{MovieLens} data, NBI can only recommend 293 distinct
objects, Hybrid-PH can recommend 787 distinct objects, while PD
increases this number to more than 1000. In \emph{Netflix} data, for
the case $L=50$, more than 5000 distinct objects can be recommended
through PD algorithm, namely almost every object has the chance to
be recommended. Secondly, the curves for NBI are remarkably steeper
than these from Hybrid-PH and PD. Take the \emph{MovieLens} data for
example (the case $L = 50$), with NBI algorithm, six movies are
recommended over six hundred times. Since there are only 943 users
in this dataset, it means that each of these movies is recommended
to more than two-thirds of the users. The result with Hybrid-PH is
much better, the No.1 object is recommended 341 times. However,
comparing with Hybrid-PH, see the insets of figure~\ref{Q_a}, PD
performs better, which indicates that with PD algorithm users are
more likely to be recommended with different objects, namely PD can
provide more personalized recommendations.

\begin{figure}
\scalebox{0.75}[0.75]{\includegraphics{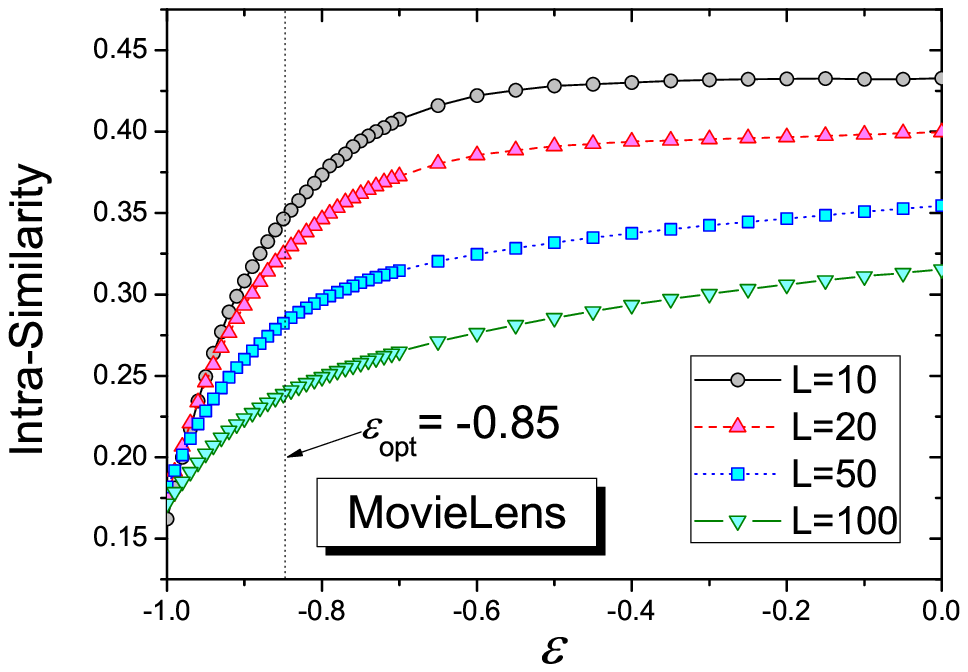}}
\scalebox{0.75}[0.75]{\includegraphics{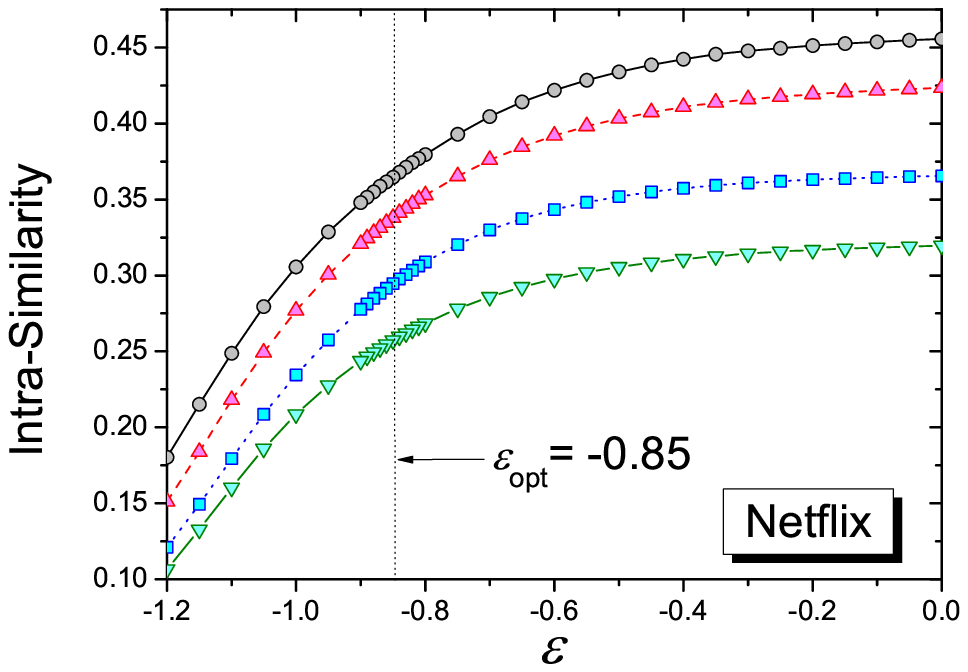}}\caption{(Color online) The
intra-similarity as a function of $\varepsilon$. Each data point is
obtained by averaging over five independent runs with data division
identical to the case shown in figure~\ref{RS_a}. The vertical
dotted line indicates the optimal parameter $\varepsilon$ subject to
the lowest ranking score.}\label{IntraS_a}
\end{figure}

\begin{figure}
\scalebox{0.65}[0.65]{\includegraphics{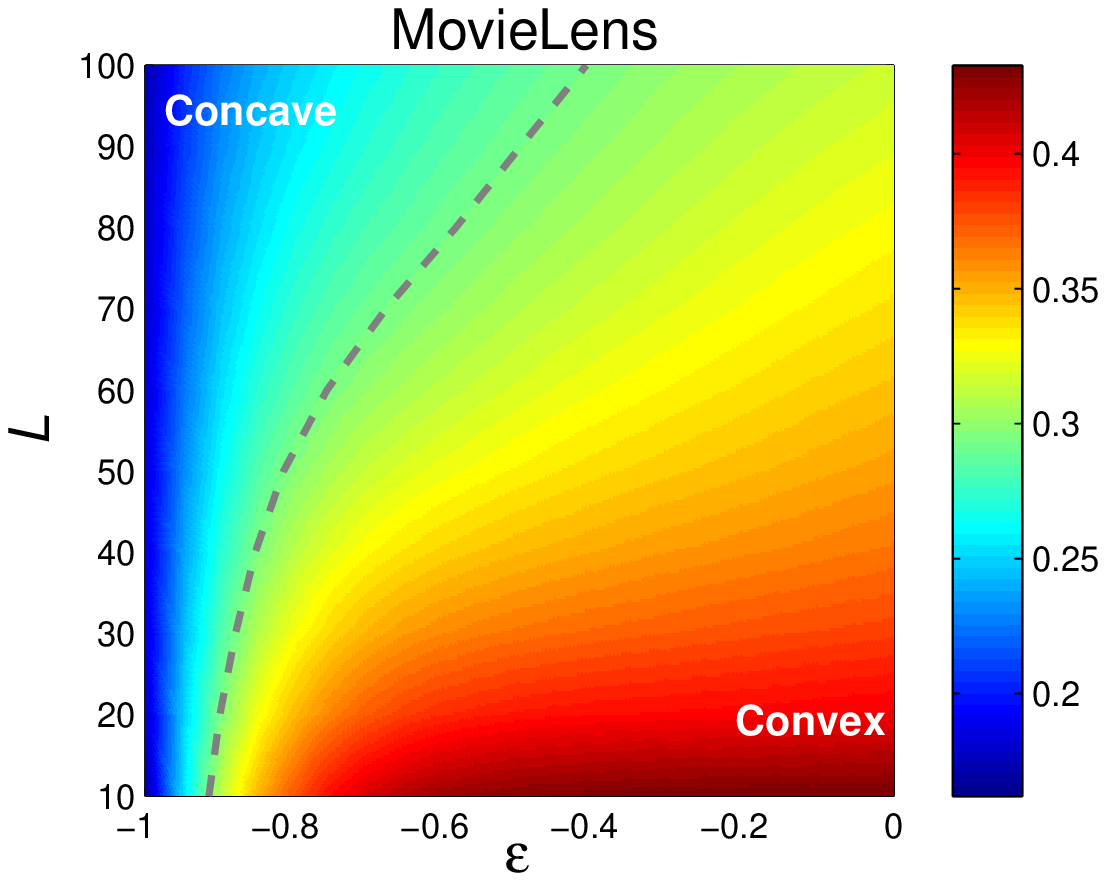}}
\scalebox{0.65}[0.65]{\includegraphics{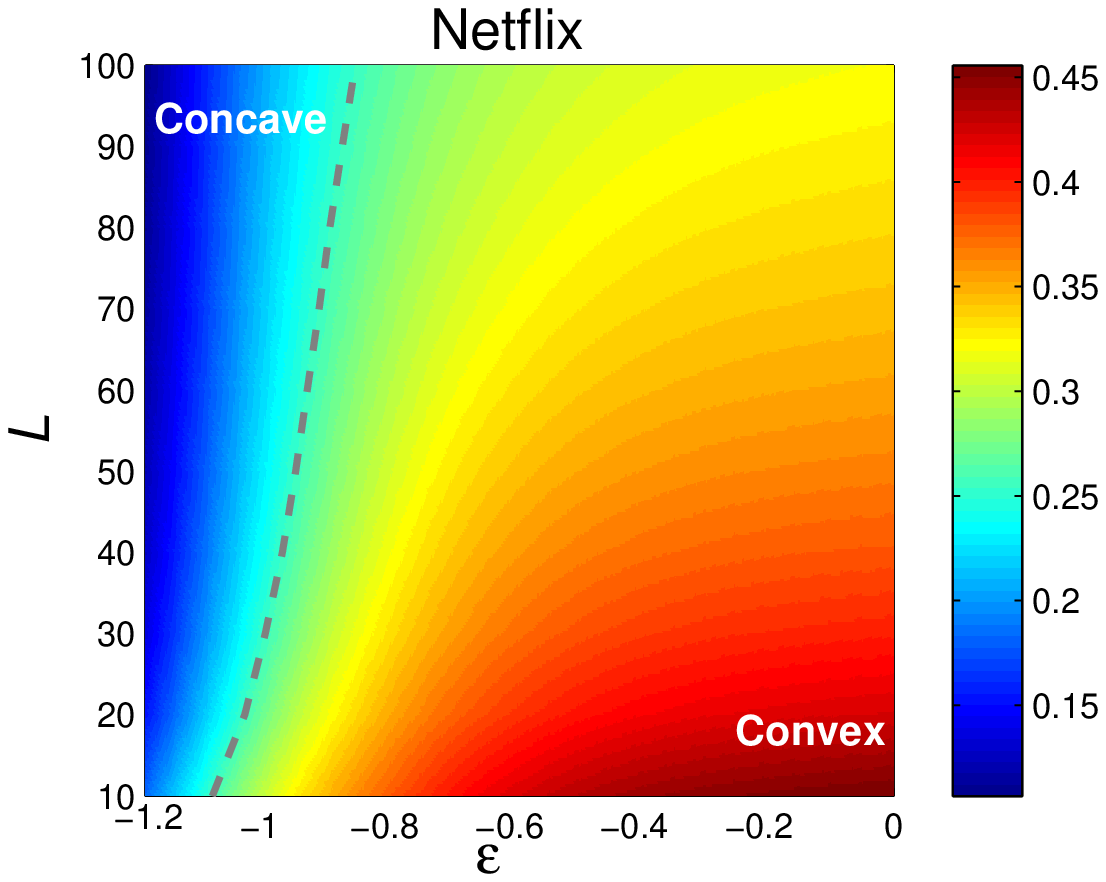}} \caption{(Color online) The
intra-similarity $I$ in ($\varepsilon$, $L$) plane for two datastes.
The numerical simulation run over the parameter $L$ in the interval
[10,100] with step length equal to 10, and the parameter
$\varepsilon$ in the interval [-1,0] and [-1.2,0] for
\emph{MovieLens} and \emph{Netflix} respectively, with step 0.05.
All the results are obtained by averaging over five independent runs
with data division identical to the case shown in figure~\ref{RS_a}.
The dashed line indicates that with the parameter combination
($\varepsilon$, $L$) on this line the intra-similarity equals the
value of the system.}\label{temp}
\end{figure}

Another metric to measure the algorithm's diversity is
intra-similarity. Different from Hamming distance, intra-similarity
measures the ability that an algorithm provides diverse
recommendations for a single user. The dependence of
intra-similarity on parameter $\varepsilon$ is shown in
figure~\ref{IntraS_a}. It shows that the parameter $\varepsilon$ is
positively correlated with intra-similarity, namely the smaller
$\varepsilon$ the lower intra-similarity (i.e., higher
intra-diversity). Comparing with NBI, when $L=50$, intra-similarity
can be decreased by 21\% for \emph{MovieLens} and 23\% for
\emph{Netflix} with optimal parameters corresponding to their
respective lowest ranking scores. Even comparing with the Hybrid-PH
algorithm, the improvement can reach up to 5\% for both datasets.
This claims that our method is effective to generate more fruitful
recommendations. Furthermore, we investigate how the two parameters
($\varepsilon$, $L$) affect intra-similarity. The intra-similarity
$I$ in ($\varepsilon$, $L$) plane for two datasets are shown in
figure~\ref{temp}. The dashed line indicates the intra-similarity of
the system which is obtained by averaging $s_{\alpha\beta}^o$ over
all the object pairs. Thus the intra-similarity as obtained from
($\varepsilon$, $L$) on the dashed line is equal to that of $L$
randomly chosen objects from the system. The left region has lower
intra-similarity while the right region has higher intra-similarity.
As a metaphor, one can think the dashed line as a plane lens keeping
the same size of the user's vision. And in the left region
especially the area corresponding to smaller $\varepsilon$ and
larger $L$, the algorithm is like a concave lens that broadens the user's
vision, while in the right region corresponding to larger
$\varepsilon$ and smaller $L$, the algorithm is like a convex lens that
narrows user's vision. The focal length is determined by parameter
$\varepsilon$. A smaller $\varepsilon$ in the left region indicates
a smaller focal length for concave lens, and hence a broader view, while in the right region indicates a larger
focal length for convex lens, hence a narrow view.

\begin{figure}
\scalebox{0.75}[0.75]{\includegraphics{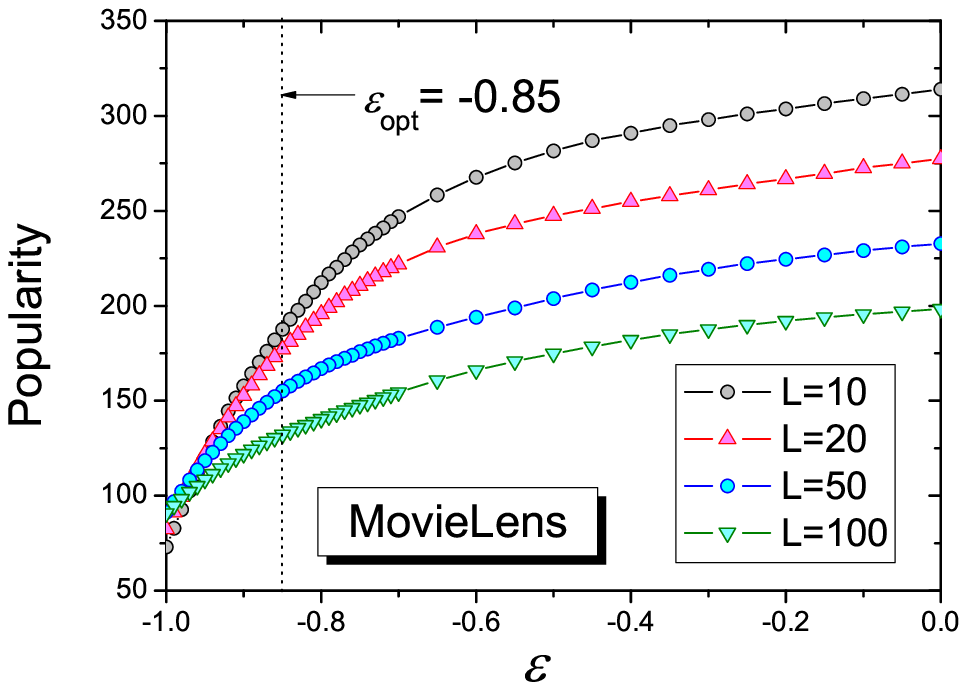}}
\scalebox{0.75}[0.75]{\includegraphics{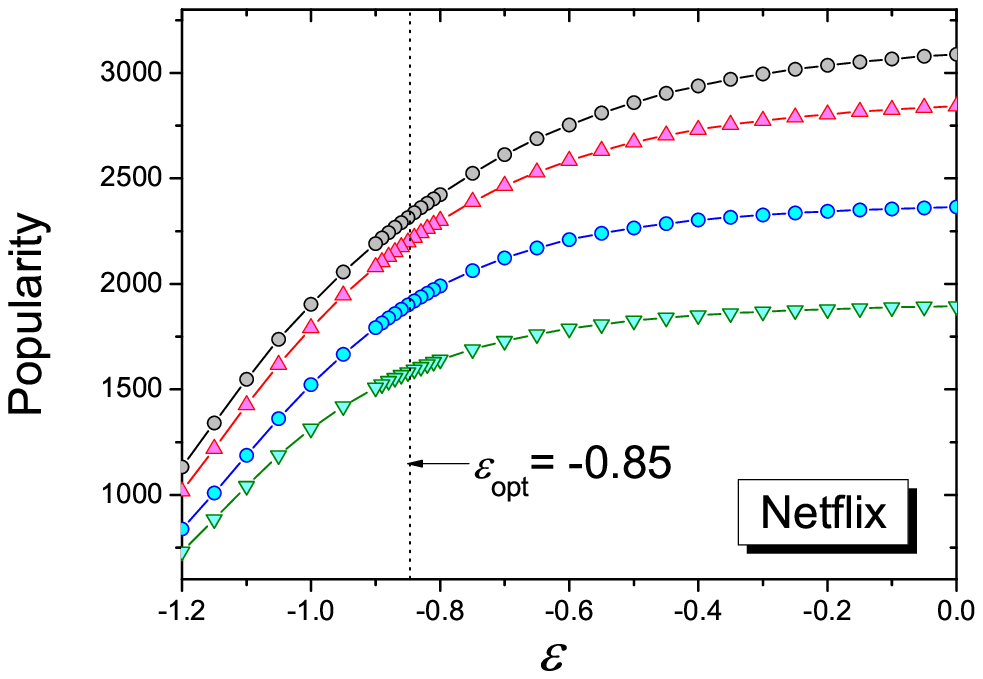}}\caption{(Color online) The
dependence of popularity (i.e., average degree) on $\varepsilon$.
Each data point is obtained by averaging over five independent runs
with data division identical to the case shown in figure~\ref{RS_a}.
The vertical dotted line indicates the optimal parameter
$\varepsilon$ subject to the lowest ranking score.}\label{Pop_a}
\end{figure}

In figure~\ref{Pop_a}, we report the dependence of popularity on
parameter $\varepsilon$. Similar with intra-similarity, a smaller
$\varepsilon$ yields a smaller popularity $P$, and thus a more novel
recommendation. Comparing with the NBI, popularity can be remarkably
improved by 33\% and 23\% for \emph{MovieLens} and \emph{Netflix}
datasets. Even comparing with the Hybrid-PH algorithm, the
improvement can reach 7\% and 5\% respectively.

\section{Effects of data sparsity}
\begin{figure}
\scalebox{0.3}[0.3]{\includegraphics{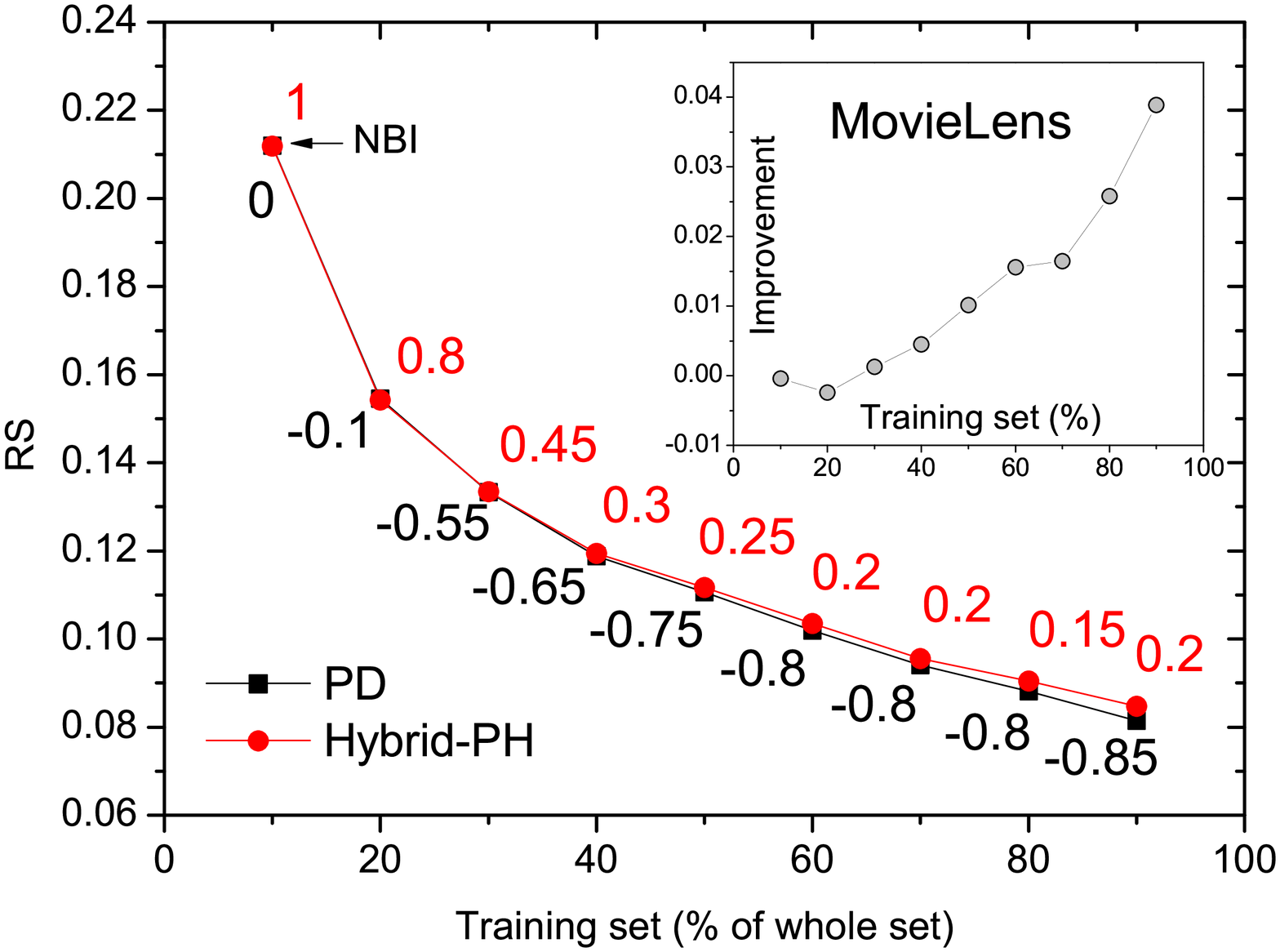}}
\scalebox{0.3}[0.3]{\includegraphics{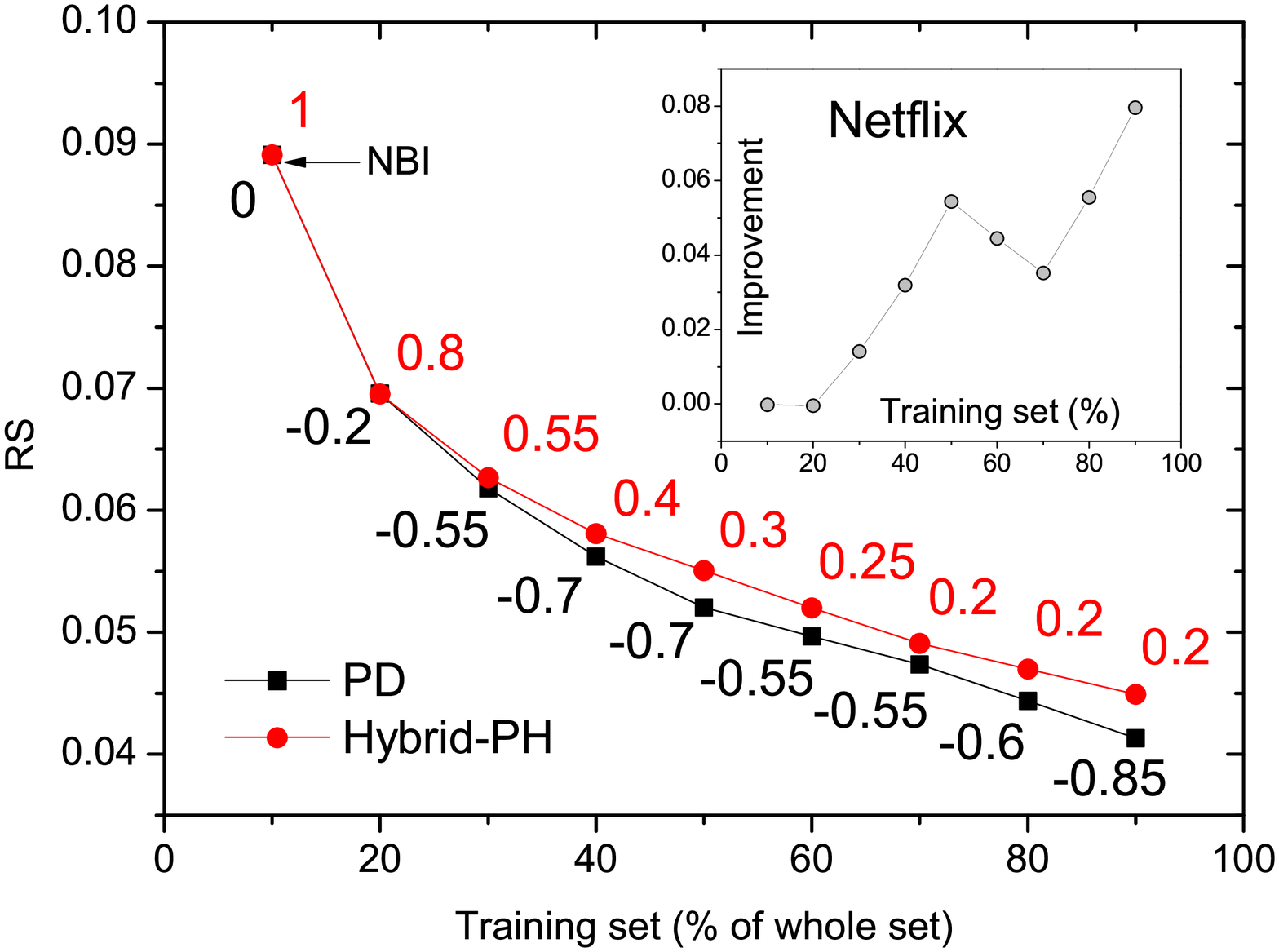}}\caption{(Color online) The
ranking score changes with the size of training set measured by the
percentage of the whole data set. That is to say, we change the size
of training set from 10\% to 90\% to respectively predict the rest
90\% to 10\%. Each data point is obtained with the parameter
($\varepsilon\in[-1,0]$ for PD and $\lambda\in[0,1]$ for Hybrid-PH
with step 0.05) subject to the lowest \emph{RS}. The optimal
parameters are labelled in black for PD and red for Hybrid-PH.
Insets show the \emph{RS}-improvement of PD comparing with Hybrid-PH
against the size of training set.}\label{sparse1}
\end{figure}

\begin{figure}
\scalebox{0.3}[0.3]{\includegraphics{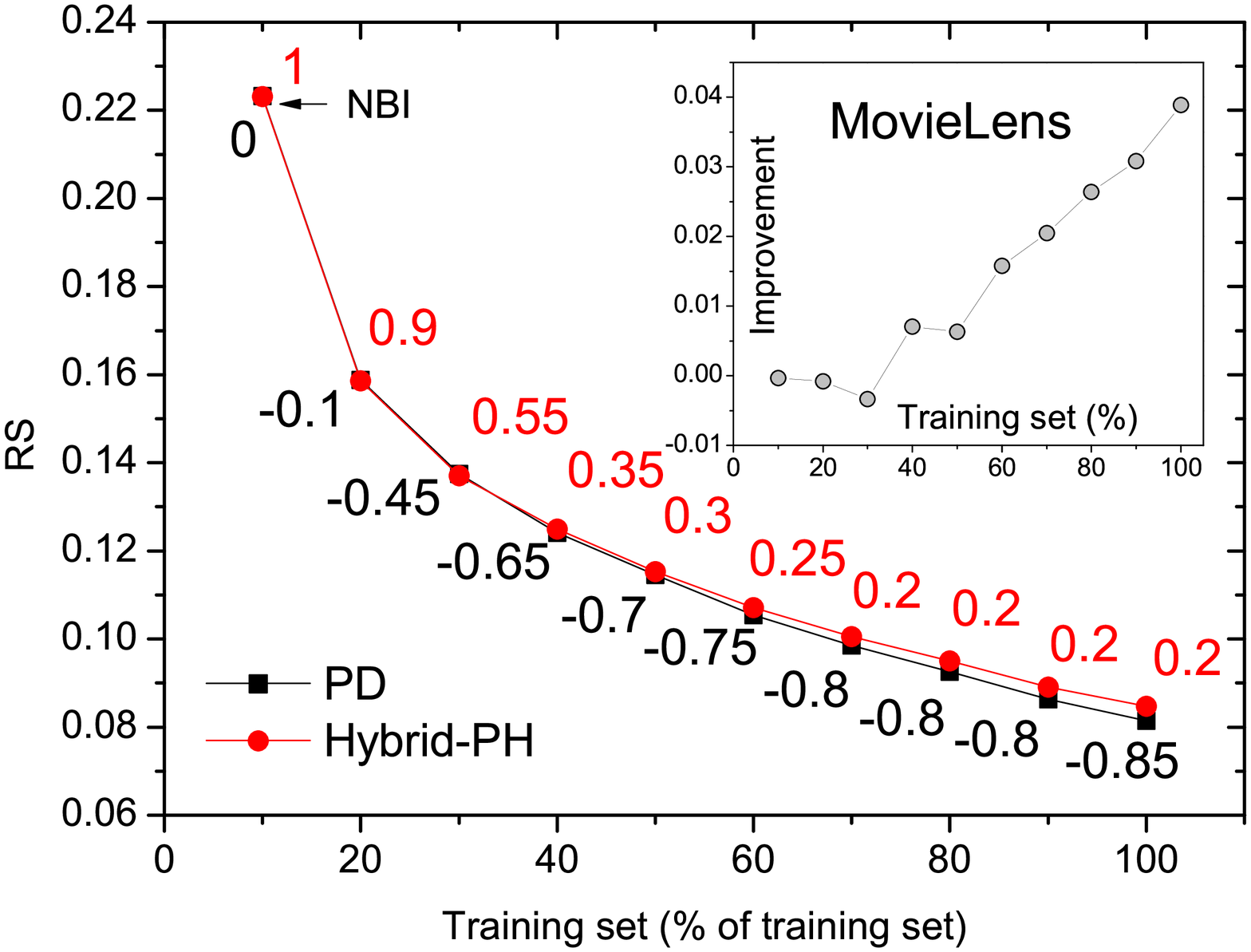}}
\scalebox{0.3}[0.3]{\includegraphics{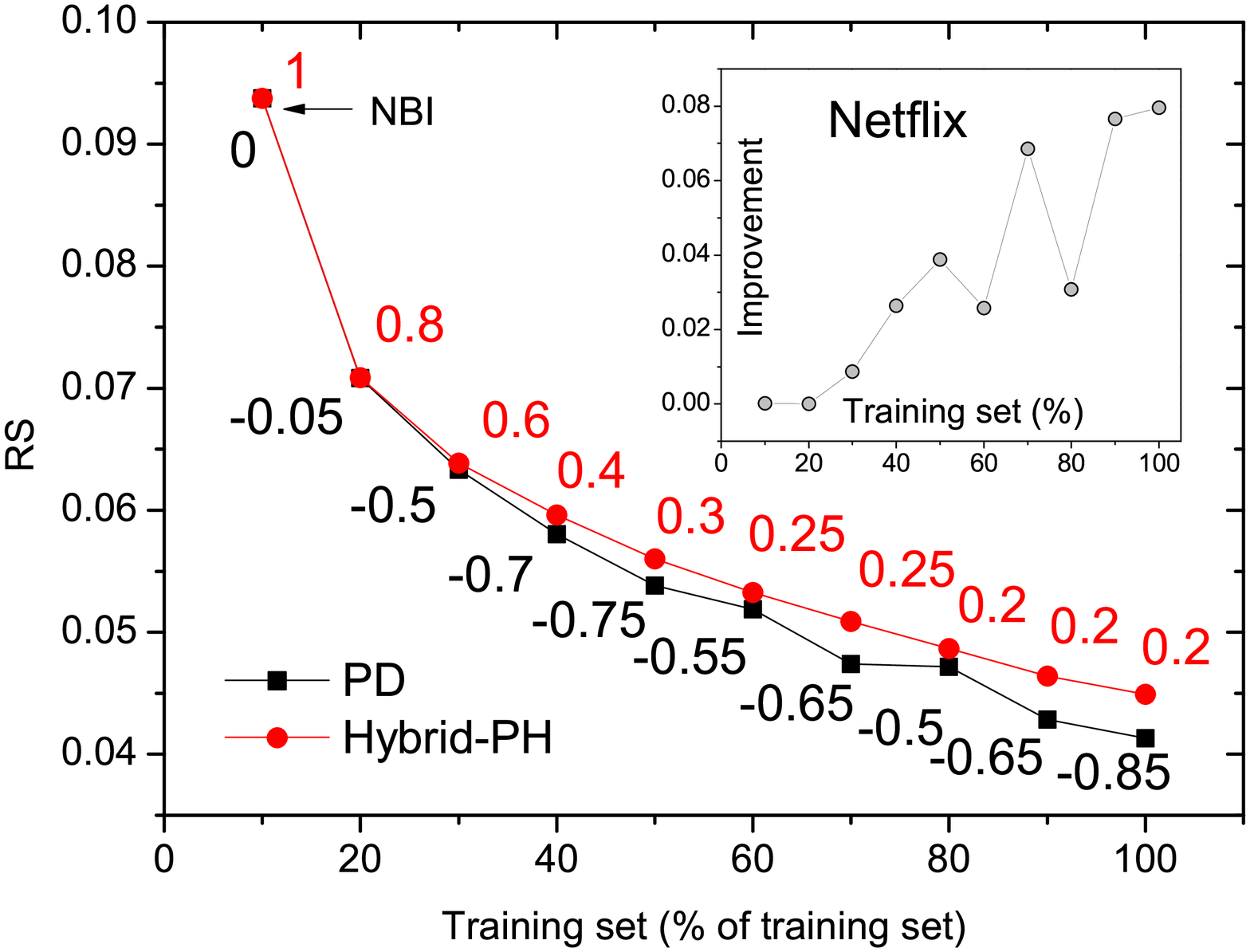}}\caption{(Color online) The
ranking score changes with the size of training set measured by the
percentage of the 90\% training set. That is to say, given a
90\%-10\% division of training set and probe set, we randomly choose
$p$\% of the known links in training set to predict the links in the
unchanged probe set. Each data point is obtained with the parameter
($\varepsilon\in[-1,0]$ for PD and $\lambda\in[0,1]$ for Hybrid-PH
with step 0.05) subject to the lowest \emph{RS}. The optimal
parameters are labelled in black for PD and red for Hybrid-PH.
Insets show the \emph{RS}-improvement of PD comparing with Hybrid-PH
against the size of training set.}\label{sparse2}
\end{figure}

In this section, we investigate the effects of data sparsity on the
algorithmic performance. Since Hybrid-PH is the most similar
algorithm with our method, we choose it for comparison (although
RE-NBI is more accurate, it considers the high-order correlations
between objects). We investigate the effects of data sparsity on the
algorithmic performance in two ways: (i) For the whole dataset, we
select $p$\% (ranging from 10\% to 90\% with step 10\%) links as
training set, and the rest $(100-p)$\% links constitute the probe
set. Clearly, lower $p$ indicates sparser data (i.e., less
information). (ii) Given a 90\%-10\% division of training set and
probe set, we randomly choose $p$\% of the known links in the
prepared training set to predict the links in probe set. To do this,
the probe links keep unchanged. For example, $p=$10 means that we
actually use 9\% of the whole dataset to predict the links in probe
set which contains 10\% links of the whole dataset. Lower $p$
indicates sparser data. The numerical results on two datasets are
shown in figure~\ref{sparse1} for method (i) and
figure~\ref{sparse2} for method (ii). Each point is obtained with
the optimal parameter subject to the lowest ranking score. From
figure~\ref{sparse1}, it can be seen that the ranking score
decreases with the increasing of the size of the training set, which
agrees with the intuition that we can obtain better recommendation
with more information. Furthermore the optimal parameters of both
methods decrease with the increasing of $p$ for both methods. It
shows that when training set contains 10\% links, the optimal
parameters are $\lambda=1$ for Hybrid-PH and $\varepsilon=0$ for PD,
which are all corresponding to the standard case NBI. Insets show
the RS-improvement of PD comparing with Hybrid-PH, which is defined
as
\begin{equation}
Improvement=\frac{RS_{Hybrid}^*-RS_{PD}^*}{RS_{Hybrid}^*},
\end{equation}
where $RS^*$ indicates the lowest ranking score for a given training
and probe set division. Generally speaking, the
\emph{RS}-improvement increases with the increasing of the size of
training set. That is to say, PD performs much better than Hybrid-PH
for denser datasets. The qualitative behaviors in
figure~\ref{sparse2} are the same as what we obtained in
figure~\ref{sparse1}, which further demonstrates that PD can give
much better predictions than Hybrid-PH for denser datasets.

\section{Conclusion and Discussion}
The preferential diffusion proposed in this paper is a kind of
biased random walk taking into account the heterogeneity of users'
degrees. The present process indeed defines a new local index of
similarity in bipartite networks (like the original NBI algorithm is
corresponding to the so-called resource-allocation similarity index
\cite{Zhou2009,Lu2009}) and thus it has potential applications in
similarity-based link prediction \cite{Liben-Nowell2007,Lu2011},
community detection \cite{Pan2010}, node classification
\cite{ZhangQM2010}, and so on. The biased random walk itself has
already found extensive application in many branches of science and
engineering, including detecting the navigation rules on complex
network \cite{Fronczak2009}, quantifying the centrality of vertex
and edge \cite{Lee2009}, modeling the animal movements
\cite{Codling2010} and information discovery in wireless sensor
networks \cite{Rachuri2009}. Here we applied the biased random walk
in dealing with the information filtering process, which may also
broaden the understanding of the applicability of biased random walk

Accuracy metrics have been widely used to evaluate the performance
of recommendation algorithms and considered to be the most important
factor. For example, the Netflix Prize \cite{Bennett2007} focuses
only on accuracy. However, user satisfaction is not always
correlated with high recommendation accuracy
\cite{Ziegler2005,McNee2002}. The recommendations on popular objects
(those are more easily to be found in other channels) are less
likely to excite users. On the contrary, the unexpected and
fortuitous recommendations which are usually related with cold
objects are more favorable. Such serendipity recommendation will
improve user experience and thus enhance their loyalty to the
system. In order to provide accurate as well as diverse and novel
recommendations, in this paper, motivated by the perspective of
physics, we proposed an algorithm, named PD, based on preferential
diffusion process on bipartite networks. We tested our algorithm on
two benchmark datasets, \emph{MovieLens} and \emph{Netflix}, and
applied five metrics, from the aspects of accuracy, diversity and
novelty, to evaluate the algorithmic performance. Comparing with the
standard algorithm NBI, the accuracy measured by ranking score can
be further improved by 23\% for \emph{MovieLens} and 18\% for
\emph{Netflix}. Even comparing with the state-of-the-art algorithm,
Hybrid-PH, the improvement can reach 4\% for \emph{MovieLens} and
9\% for \emph{Netflix}. Moreover, the performance of PD can be
further improved by considering a heterogenous initial resource
configuration.

Furthermore, statistical result on the ranking score of individual
objects shows that our method has much higher ability to accurately
recommend the low-degree objects. That is to say, such prominent
improvement on accuracy comes mainly from the highly accurate
recommendation on unpopular objects, and thus it indeed enhances the
recommendation diversity and novelty. For example, if we recommend
50 objects to each user, in \emph{MovieLens}, NBI can only recommend
293 distinct objects to all users, Hybrid-PH can recommend 787
distinct objects, while PD increases this number to more than 1000.
In \emph{Netflix} data, more than 5000 distinct objects can be
recommended through PD algorithm, namely almost every object has the
chance to be recommended. Specially, we found that the recommender
system may play different roles from the aspect of intra-similarity
--- the similarity within a user's recommendation list, which is
determined by the algorithm's parameter $\varepsilon$ and the length
of recommendation list $L$. Given ($\varepsilon$, $L$), if the
intra-similarity generated by algorithm is higher than that of $L$
randomly selected objects (i.e., average intra-similarity of the
whole system), the recommender system plays the role as a convex
that narrows users' vision, whereas if intra-similarity generated by
algorithm is lower than that of the system, the recommender system
plays the role as a concave that broadens users' vision. Besides, we
investigated the dependence of algorithm performance on data
density. The results show that comparing with Hybrid-PH, PD
algorithm gives more significant improvement for denser data.

A good recommendation algorithm can guide the system for a better
development. You can think that the system itself and the
recommendation algorithm constitute a symbiotic system. Generally
speaking, there is no best recommendation algorithm, but the most
suitable algorithm for a given system or a user. Just like the
marriage game \cite{Omero1997}: choose the right but not the best.
In this sense, the most equitable evaluation on recommendation
algorithm should be based on the user experience which is difficult
to capture in metric. Notice that, the optimal algorithm (or
parameter) for the whole system is usually different from the
optimal algorithm (or parameter) for an individual user. Thus an
applicable and feasible way is building an open recommender system
where users can help themselves to find their best experienced
algorithm (or parameter). For example, we can set a bar controlling
the parameter of the algorithm on the website. Take the PD algorithm
as an example, the user may set large value of $\varepsilon$ to obtain recommendations of popular and hot items, and set small value of $\varepsilon$ to obtain recommendations of niche and novel items.
Here we argue that the design of user-centric recommender systems
will become one of the challenges of the next generation information
filtering techniques. Finally, we believe that this paper may shed
some light on this interesting and exciting direction.


\section*{Acknowledgments}
We acknowledge the \emph{GroupLens Research Group} for
\emph{MovieLens} data and the \emph{Netflix Inc.} for \emph{Netflix}
data. We thank Yi-Cheng Zhang for providing the proper metaphor of
\emph{Concave} and \emph{Convex} when referring the user
intra-similarity, and Tao Zhou for a critical reading of the
manuscript. This work is partially supported by the Swiss National
Science Foundation under Grant No. (200020-132253) and the National
Natural Science Foundation of China under Grant Nos. (60973069,
90924011,11075031).

\end{document}